\UseRawInputEncoding

\documentclass[aps,prx,amsmath,amssymb,reprint,floatfix]{revtex4-2}

\usepackage{hyperref}
\usepackage{graphicx}
\graphicspath{{images/}}
\usepackage{dcolumn}
\usepackage[utf8]{inputenc}
\usepackage{amssymb}
\usepackage{graphicx}
\usepackage[dvipsnames]{xcolor}
\usepackage{amsmath}
\usepackage{bm}
\usepackage{float}

\begin{document}

\title{Hybrid SU(1,1) interferometry in optomechanics}

\author{Chao Meng}
\author{Emil Zeuthen}
\email{zeuthen@nbi.ku.dk}
\affiliation{Niels Bohr Institute, University of Copenhagen, Blegdamsvej 17, 2100 Copenhagen \O{}, Denmark}
\author{Polina R. Sharapova}
\email{polinash@campus.uni-paderborn.de}
\affiliation{    Department of Physics, Paderborn University,
        Warburger Straße 100, D-33098 Paderborn, Germany}

\date{\today}

\begin{abstract} 
In non-degenerate SU(1,1) interferometers, beam splitters are replaced by two-mode squeezers, enabling sub-shot-noise sensitivity without input squeezing and robustness to detection losses by quantum entanglement. 
We propose a hybrid implementation in optomechanics where one ``arm'' is a mechanical mode undergoing two consecutive, mode-matched interactions with a traveling optical field (constituting the other arm). 
Such engineered interactions allow for sub-shot-noise phase detection even in the presence of mechanical thermal noise and optical losses, advancing precision interferometry in hybrid systems.

\end{abstract}

\maketitle

Interferometry is central to precision measurement and underpins applications from sensing and imaging to fundamental tests of physics. Optomechanical  interferometers \cite{aspelmeyer_cavity_2014, bowen_quantum_2015}, for example, enabled the first detections of gravitational waves \cite{abbott_observation_2016} and quantum states preparation~\cite{rossi_measurement-based_2018, meng_mechanical_2020, meng_measurement-based_2022}. With coherent light, phase sensitivity is bounded by the shot-noise limit (SNL). Injecting squeezed light can surpass the SNL \cite{jia_squeezing_2024,tse_quantum-enhanced_2019, vahlbruch_observation_2008}, but this approach is vulnerable to optical and detection losses. 

SU(1,1) interferometers  \cite{yurke_su2_1986,ou_quantum_2020,salykina_sensitivity_2023,chekhova_nonlinear_2016, ou_quantum_2020}, which involve two \emph{nonlinear} interactions, are robust to detection losses and can overcome the SNL with fewer optical elements compared to conventional [SU(2)] interferometry ~\cite{liang_phase_2022,salykina_sensitivity_2023,manceau_detection_2017}. Such interferometers are characterized by strong correlations between the generated signal and idler beams \cite{wang_induced_1991, lemos_quantum_2014, kalashnikov_infrared_2016,defienne_advances_2024}. 
SU(1,1) interferometers present a powerful paradigm that has already demonstrated significant advantages in numerous systems, including optical \cite{manceau_detection_2017,li_phase_2014,li_phase_2016,adhikari_phase_2018}, microwave \cite{flurin_generating_2012}, atom-atom \cite{gross_nonlinear_2010,gabbrielli_spin-mixing_2015,linnemann_quantum-enhanced_2016}, phonon-phonon \cite{patil_thermomechanical_2015}, hybrid atom-light \cite{chen_atom-light_2015,szigeti_pumped-up_2017} and integrated \cite{ferreri_spectrally_2021,ferreri_two-colour_2022} platforms. 
In particular, hybrid implementations couple a localized oscillator (e.g., mechanical) to a traveling, generally multimode optical field. The temporal modes addressed by the two interaction stages are therefore generically mismatched \cite{barakat_simultaneous_2025,chen_atom-light_2015,scharwald_phase_2023}. This mode mismatch diverts light into orthogonal modes and is equivalent to an internal loss channel, a key bottleneck for multimode SU(1,1) interferometers that degrades phase sensitivity. 

In this Letter, we address this fundamental limitation by deriving an analytical expression for the appropriate shaping of the pair of drive pulses required for perfect mode-matching (in absence of loss and decoherence). 
Then, for such an engineered pair of drive pulses, we evaluate the phase sensitivity of our scheme in the presence of the asymmetrical loss and thermal decoherence characteristic of hybrid platforms. 
We find that sub-SNL performance is achievable under realistic experimental conditions even for moderate squeezing strengths, provided that optical losses can be kept low and at most a few thermal quanta enter during the protocol (on average). 
Our work eliminates the primary bottleneck in hybrid SU(1,1) interferometry and unlocks its full potential, paving the way for applications in quantum sensing and hybrid quantum networking. 
For the sake of concreteness, we will illustrate our scheme by analyzing an optomechanical implementation. 
However, our proposal is quite general and can be applied to other types of hybrid systems, e.g., atom-light interferometers~\cite{chen_atom-light_2015}. 

\begin{figure}[t]
\includegraphics[width=\columnwidth]{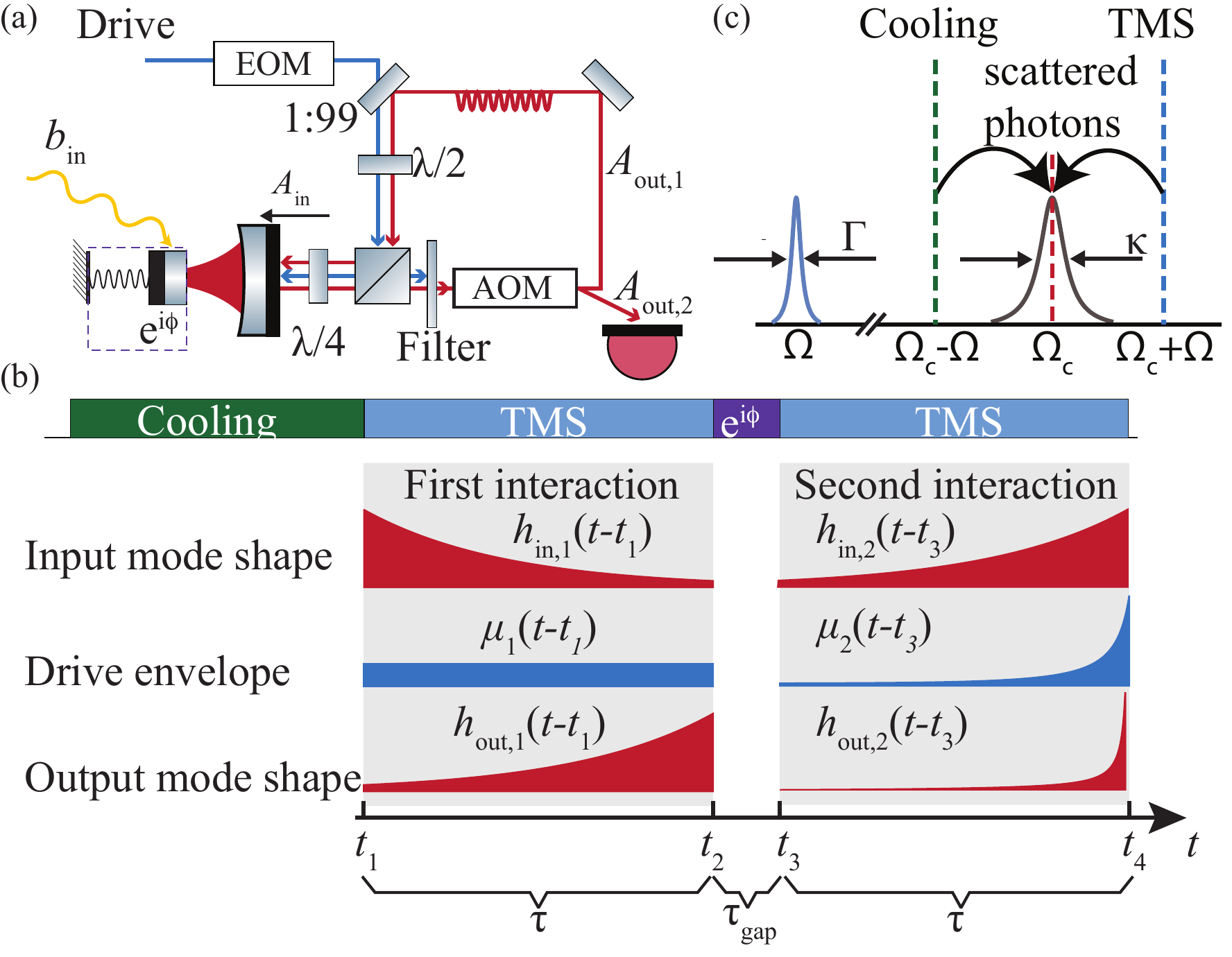}
	\caption{Optomechanical SU(1,1) interferometer. (a) Sketch of the proposed setup. Pulsed driving light (blue) is transformed by an electro-optical modulator (EOM) to obtain the required temporal profile and then enters via the highly reflective mirror (labelled 1:99). A polarizing beam splitter (PBS) couples the driving light into the cavity where the TMS-type nonlinear optomechanical interaction occurs. The generated TMS light (red) is then directed into a delay loop while the driving light is blocked by a narrow-band optical filter. The half-wave and quarter-wave plates serve to align the polarization between $\hat{A}_{\mathrm{in},2}$ and $\hat{A}_{\mathrm{out},1}$. For the second interaction, a new pulse of driving light is shaped according to the rising profile presented in (b). The acousto-optic modulator (AOM) is activated for the second interaction, to route TMS photons to the detector.   
(b) The time protocol illustrating the shapes of the light modes participating in the optomechanical interaction for a particular pair of drive-light pulses (separated by $\tau_\mathrm{gap}$) that achieve mode matching, i.e., $h_{\mathrm{out},1}(t') = h_{\mathrm{in},2}(t')$. In this figure, $t\in[t_1,t_4]$ denotes the absolute time of the protocol, whereas mode and drive envelopes $h_{x,j}(t')$ and $\mu_j(t')$, $t'\in[0,\tau]$, are referenced to the beginning of their interaction interval $j$.
 (c) Implementation of the TMS-type optomechanical interaction using the blue-detuned drive light $\Omega_L=\Omega_c+\Omega$ that results in generation of cavity photons at $\Omega_c$ (black lineshape, full-width-at-half-maximum (FWHM) $\kappa$) entangled with the phonons at $\Omega$ (blue lineshape, FWHM $\Gamma\ll\kappa$); optionally, precooling can be achieved by red-detuned driving $\Omega_L=\Omega_c-\Omega$.} 
 \label{fig:Sketch} 
\end{figure}

\emph{Optomechanical implementation}---The protocol is sketched in Fig.~\ref{fig:Sketch} and is based on a one-sided, resolved-sideband optomechanical cavity undergoing a sequence of two pulsed  interactions~\cite{palomaki_entangling_2013,fiaschi_optomechanical_2021} of the two-mode squeezing (TMS)~\cite{aspelmeyer_cavity_2014,bowen_quantum_2015} type [Fig.~\ref{fig:Sketch}(a)]. 
The purpose of these two interactions is to squeeze ($t=t_1$ to $t_2$) and anti-squeeze \cite{caves_reframing_2020}
($t=t_3$ to $t_4$) the optomechanical modes; however an unknown mechanical phase shift $\phi$ is applied between the two interactions ($t=t_2$ to $t_3$), altering the interference. 
Whereas the optical input fluctuations for the first TMS interaction are assumed to be those of the vacuum, the generated output light field loops back via the delay line into the optomechanical system to serve as the optical input fluctuations for the second interaction. To match the input and output temporal modes, $h_{\mathrm{out},1}(t')=h_{\mathrm{in},2}(t')$ is ensured by appropriate shaping of the drive-pulse envelopes [Fig.~\ref{fig:Sketch}(b)]. The resulting output light mode $\hat{A}_{\mathrm{out},2}$ can be highly sensitive to the signal phase $\delta\phi$ as can be extracted from a photon counting measurement.

The TMS interaction is readily achieved in a resolved-sideband ($\kappa \ll \Omega$) optomechanical system by blue-detuned driving at laser frequency $\Omega_L=\Omega_c+\Omega$, where $\Omega_c$ and $\Omega$ are the cavity and mechanical resonances, respectively, and $\kappa$ is the cavity decay rate. In the regime of strong driving, this 
results in the generation of correlated photons with central frequency $\Omega_c$ and phonons at $\Omega$ [Fig.~\ref{fig:Sketch}(c)], corresponding to interaction Hamiltonian $\hat{H}_\mathrm{int}=\hbar g(t)(\hat{a}^{\dagger} \hat{b}^{\dagger}+\hat{a} \hat{b})$ \cite{law_interaction_1995}, 
where $g(t)$ is the interaction strength proportional to the instantaneous coherent drive amplitude in the cavity $\alpha(t)$~\footnote{The coherent drive also induces a time-dependent, deterministic mean displacement to the mechanical system, which can be compensated by additional actuation and therefore is neglected here.}, while $\hat{a}^{\dagger}$ ($\hat{a}$) and $\hat{b}^{\dagger}$ ($\hat{b}$) are the zero-mean creation (annihilation) operators of the intracavity photon and phonon fluctuations, respectively~\cite{aspelmeyer_cavity_2014}. 

The corresponding Heisenberg-Langevin equations of motion are, after including coupling to extracavity optical modes and the mechanical bath \cite{aspelmeyer_cavity_2014,bowen_quantum_2015}, 
\begin{subequations}\label{eq:eq_motion}
	\begin{align}
	 \dot{\hat{a}}(t)&=-i g(t) \hat{b}^{\dagger}(t)-\frac{\kappa}{2} \hat{a}(t)+\sqrt{\kappa} \hat{a}_{\mathrm{in}}(t) \\
	\dot{\hat{b}}(t)&=-i g(t) \hat{a}^{\dagger}(t)-\frac{\Gamma}{2} \hat{b}(t)+\sqrt{\Gamma} \hat{b}_{\mathrm{in}}(t), \label{eq:eq_motion-b}
 	\end{align}
\end{subequations}
where $\Gamma$ is the mechanical damping rate, $\hat{a}_{\mathrm{in}}$ is the cavity input field, and $\hat{b}_{\mathrm{in}}$ is the input field of the mechanical bath, which we suppose to be Markovian with occupancy $n_\mathrm{th}$,  
$\langle\hat{b}_\mathrm{in}^\dagger(t) \hat{b}_\mathrm{in}(t') \rangle=n_\mathrm{th}\delta(t-t')$. 
We assume a high mechanical quality factor such that $\Gamma \ll \Omega,1/(t_4-t_1)$, whereby intrinsic mechanical damping becomes negligible whereas the decoherence rate $n_{\mathrm{th}}\Gamma$ may be significant. 
The cavity output field is given by the input-output relation $\hat{a}_\mathrm{out}(t)=\hat{a}_\mathrm{in}(t)-\sqrt{\kappa}\hat{a}(t)$. 
Under a weak optomechanical coupling ($g\ll\kappa$) considered henceforth, one can adiabatically eliminate the cavity mode,  
resulting in the effective optomechanical input-output relation $\hat{a}_\mathrm{out}(t)=-\hat{a}_\mathrm{in}(t)+i\sqrt{\mu(t)}\hat{b}^\dagger(t)$, where we have introduced the interaction rate between the mechanical mode and the traveling field $\mu(t)=4g(t)^2/\kappa$.

\emph{Two-mode limit (negligible thermal decoherence)}---The dynamics~\eqref{eq:eq_motion} leads to ideal TMS interaction between the external light field and the mechanical mode only insofar as we can neglect the thermal noise term $\propto\hat{b}_\mathrm{in}$ in Eq.~\eqref{eq:eq_motion-b}; this is the case if the duration of the TMS sequence is much shorter than the decoherence time, $t_4-t_1 \ll 1/(n_{\mathrm{th}}\Gamma)$ \cite{hofer_quantum_2011}. 
In this case, the interaction truly involves only two modes (i.e., two input modes transform to two output modes): the localized mechanical mode, $\hat{B}_{\text{in}}= \hat{b}(0)\rightarrow\hat{B}_{\text{out}}=\hat{b}(\tau)$, and 
temporal light mode,
$\hat{A}_\mathrm{in} \rightarrow \hat{A}_\mathrm{out}$ \cite{hofer_quantum_2011}, where $\hat{A}_x=\int_0^\tau h_x(t)\hat{a}_x(t) dt$ [$x\in\{\text{in},\text{out}\}$] are defined by envelopes $h_x(t)$ determined by $\mu(t)$ (see SM Sec.~\ref{app:OM-EOM-IO}),
\begin{subequations}\label{eq:2-mode-modes}
	\begin{align}
	 \hat{A}_{\text {in}}&=\sqrt{\frac{1}{1-e^{-M}}} \int_0^\tau \sqrt{\mu(t)}e^{-\frac{M(t)}{2}} \hat{a}_{\text{in}}(t) d t, \label{eq:A-in_def}\\
	 \hat{A}_{\text {out}}&=\sqrt{\frac{1}{e^M-1}} \int_0^\tau \sqrt{\mu(t)} e^{\frac{M(t)}{2}} \hat{a}_{\text{out}}(t) d t. \label{eq:A-out_def}
	\end{align}
\end{subequations}
Here, $M(t)=\int_0^t \mu(t') dt'$ is the squeezing strength of the process up to time $t$, whereas the strength of the full interaction is denoted by $M\equiv M(\tau)$. 
These operators obey the bosonic commutation relations $ [\hat{A}_x, \hat{A}_x^{\dagger}]=[\hat{B}_x, \hat{B}_x^{\dagger}]=1$, 
and are linked by the standard TMS transformation
\begin{equation}\label{eq:2-mode-IO}
	\begin{aligned}
	 \hat{A}_{\text{out}}&=-\sqrt{e^{{M}}} \hat{A}_{\text{in}}+i \sqrt{e^{M}-1} \hat{B}_{\text {in}}^{\dagger}, \\
	 \hat{B}_{\text{out}}^{\dagger}&=\sqrt{e^{{M}}} \hat{B}_{\text{in}}^{\dagger}+i \sqrt{e^{M}-1} \hat{A}_{\text {in}},
	\end{aligned}
\end{equation}
while all other temporal modes of the input light field $\hat{A}^{(j)}_{\mathrm{in}\perp}$ orthogonal to the one defining $\hat{A}_\mathrm{in}$, map passively to modes of the output field 
$\hat{A}^{(j)}_{\mathrm{out}\perp}=-\hat{A}^{(j)}_{\mathrm{in}\perp}$ 
orthogonal to $\hat{A}_\mathrm{out}$.

\emph{Mode matching of concatenated interactions}---The implementation of the SU(1,1) scheme relies on two successive TMS transformations~(\ref{eq:2-mode-IO})   
whose optical modes are properly linked: the output mode $\hat{A}_{\mathrm{out},1}$ from the first TMS stage should serve as the input mode $\hat{A}_{\mathrm{in},2}=\hat{A}_{\mathrm{out},1}$ for the second stage. 
In between the interactions, the mechanical mode $\hat{B}_{\mathrm{out},1}$ experiences a phase shift $\phi$, whereby $\hat{B}_{\mathrm{in},2} = \hat{B}_{\mathrm{out},1} e^{i\phi}$ for negligible mechanical damping.

SU(1,1) interferometers are particularly vulnerable to deviation from $\hat{A}_{\mathrm{in},2} = \hat{A}_{\mathrm{out},1}$, as quantified by the \emph{internal} efficiency $\eta_{12}$ of the interferometer,
\begin{equation}\label{eq:A-in2}
\hat{A}_{\text{in},2}=\sqrt{\eta_{12}} \hat{A}_{\text{out},1}+\sqrt{1-\eta_{12}} \hat{A}_{\text{int}}.
\end{equation}
We decompose $\eta_{12} = \eta_{\mathrm{mode}} \eta_{\mathrm{tech}}$ into contributions from: 1) temporal mode overlap $\left.\eta_{\mathrm{mode}}=\int_0^\tau h_{\mathrm{out},1}(t)h_{\mathrm{in},2}(t)dt\right.$ [see Eqs.~\eqref{eq:2-mode-modes}] for the chosen pair of drive pulses $\left.\{\mu_1(t),\mu_2(t)\}\right.$, and 2) technical efficiency $\eta_{\mathrm{tech}}$ caused by fiber-coupling and optical losses in the path storing the optical output of the first interaction [Fig.~\ref{fig:Sketch}(a)]. In the regime of negligible thermal decoherence, the added noise operator $\hat{A}_{\mathrm{int}}$ represents vacuum input. 

However, the temporal-mode contribution  can be rendered unity, $\eta_{\mathrm{mode}}=1$, by virtue of the time-variable drive field envelopes $\mu_j(t)$. 
As a specific demonstration of this, in this Letter we choose $\mu_1(t)=\mu_1$ constant and vary the envelope of the second interaction as
 \begin{equation}\label{eq:mu2-shape}
 \mu_2(t)=
 \frac{\mu_1}{\frac{e^{M_1}-e^{-M_2}}{1-e^{-M_2}}e^{-\mu_1 t}-1}; 
\end{equation}
this scenario is illustrated in Fig.~\ref{fig:Sketch}(b).
Note that matching can be achieved for a family of drive envelope pairs $\{\mu_1(t),\mu_2(t)\}$, see SM Sec.~\ref{app:shaping-sub}.
As a naïve benchmark, we note that for the choice of two boxcar pulses $\left.\mu_1(t)=\mu_2(t)=\mu\right.$ ($t\in[0,\tau]$), Eqs.~\eqref{eq:2-mode-modes} lead to the mode overlap of two exponentials, one increasing and one decreasing, $\eta_\mathrm{mode}=(M/2)/\sinh(M/2)$, which is detrimental in the regime of interest $M\equiv\mu\tau\gtrsim 1$ (see SM Sec.~\ref{app:without_shaping}).

\emph{Phase sensitivity}---For a direct intensity measurement of the optical output field centered at $\Omega_c$, (i.e., filtering out the drive at $\Omega_L$), the experimentally simplest measurement technique, the phase sensitivity can be evaluated according to the error propagation formula as \cite{yurke_su2_1986}
\begin{equation}\label{eq:sens-SU}
\Delta \phi=\left.\frac{\Delta \hat{N}_{\text{out},2}}{|\partial\langle \hat{N}_{\text{out},2}\rangle/\partial \phi|}\right|_{\phi=\phi_0},
\end{equation}
where $\langle \hat{N}_{\text{out},2}\rangle$ and $\Delta \hat{N}_{\text{out},2}$ are the mean and standard deviation of the number of quanta $\left.\hat{N}_{\mathrm{out},2}=\hat{A}_{\mathrm{out},2}^\dagger\hat{A}_{\mathrm{out},2}\right.$. 
We benchmark this phase sensitivity against the SNL 
$\Delta \phi_\mathrm{SNL} = 1/(2\sqrt{N_\mathrm{pr}})$, 
where $N_\mathrm{pr} $ is the mean number of quanta probing the phase change \cite{salykina_sensitivity_2023,ou_quantum_2020}, following the convention given by Anderson \textit{et al.}\ \cite{anderson_optimal_2017}. 
The SNL defines the classical bound, i.e., the best sensitivity achievable in a Mach-Zehnder interferometer probed by coherent light. 
In our case, 
    $\left.N_\mathrm{pr} = \langle\hat{B}_{\mathrm{out},1}^\dagger\hat{B}_{\mathrm{out},1}\rangle =(n_0+1)e^{M_1}-1\right.$, 
is the number of \emph{phonons} after the first interaction,
defined by the strength of the first interaction $M_1$ and the initial mean thermal phonon number $n_0\equiv\langle\hat{B}_{\mathrm{in},1}^\dagger\hat{B}_{\mathrm{in},1}\rangle$ (assuming the number of thermal phonons added \emph{during} the interaction to be negligible). 

\begin{figure*}[t]
\includegraphics[width=\textwidth]{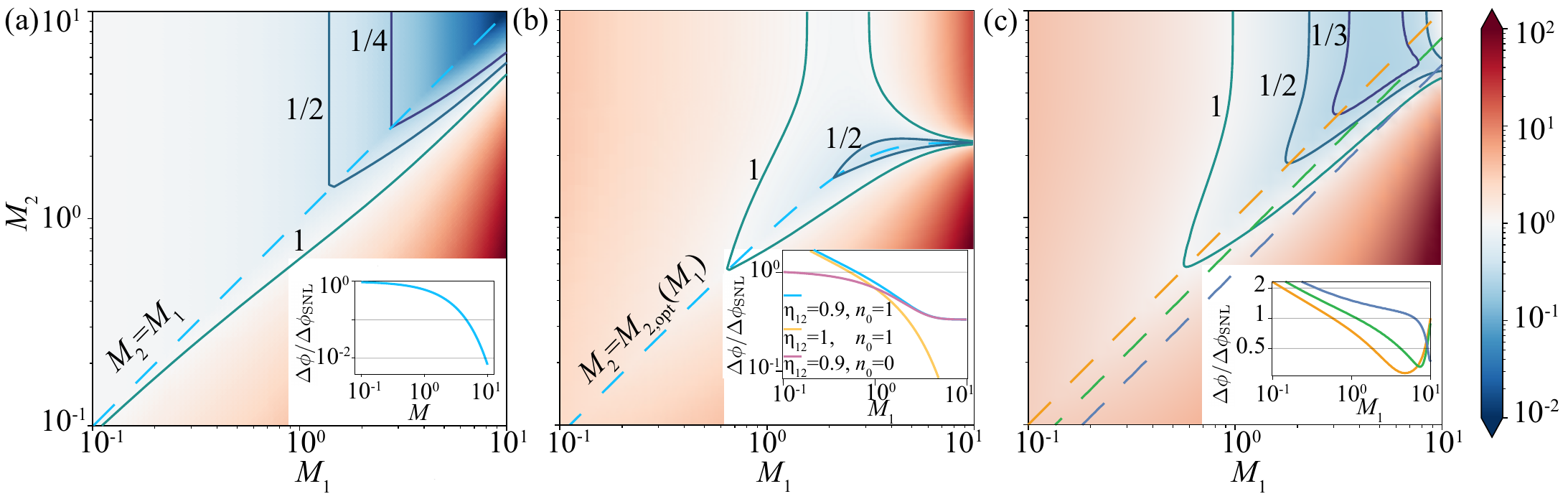}
\caption{\label{fig:ideal_case} SNL-normalized phase sensitivity $\Delta \phi/\Delta \phi_\mathrm{SNL}$. a) Idealized case with $\eta_{12}=1$ and $n_0=0$; the inset shows the sensitivity for $M_2 = M_1$ (dashed line). 
b) Lossy case with $\eta_{12}=0.9$, $\eta_{\mathrm{det}}=1$, and $n_0=1$; the inset shows the sensitivity for $M_2=M_{2,\mathrm{opt}}(M_1)$ [dashed curve, Eq.~\eqref{eq:M2-opt}] for different parameter sets. 
c) 2D photonic crystal with thermal noise at room temperature ($n_{\mathrm{th}}=6\times 10^2$) and $n_0=1$, other parameters are mentioned in the text. Imperfections include the thermal decoherence $n_\mathrm{th}\Gamma$ and delay fiber loss $\eta_\mathrm{tech}<1$. Three relations of $M_1=M_2$ (orange), $M_1=1.36M_2$ (green) and $M_1=1.86M_2$ (blue) are indicated by the dashed lines and presented in the inset.
The solid contours indicate sensitivity levels.}
 \label{fig:two-mode}
\end{figure*}
 
Fig.~\ref{fig:two-mode} presents a map of phase sensitivity, where for each given pair of squeezing strengths $(M_1, M_2)$, $\Delta \phi$ is optimized over the reference phase $\phi_0$. 
Fig.~\ref{fig:two-mode}(a) shows $\Delta \phi$ for the idealized scenario of unit internal efficiency, $\eta_{12}=1$, and zero initial mechanical occupancy, $n_0=0$, 
and reveals that efficient mode matching enables a large region of sub-SNL sensitivity for squeezing strengths $M_2 \gtrsim M_1$ (blue area). 
When either $M_1$ or $M_2$ is fixed, the minimum sensitivity $\Delta \phi |_{\phi_0=0}=1/(2\sqrt{N_\mathrm{pr}(N_\mathrm{pr}+1)})$ is achieved for $M_1 = M_2$ at $\phi_0=0$, realizing Heisenberg scaling $\sim 1/(2N_\mathrm{pr})$ \cite{manceau_improving_2017}\footnote{Ref.~\cite{manceau_improving_2017} writes the Heisenberg limit as $\sim 1/(2N)$ in terms of $N=2N_\mathrm{pr}$, the total number of quanta in the case of a balanced interferometer, while assuming the relative phase shift of the arms to be $2\phi$; taking these differences in conventions into account, one finds consistency with our work.}, see the inset in Fig.~\ref{fig:ideal_case}(a).

For finite internal efficiency $\eta_{12}<1$ [Fig.~\ref{fig:two-mode}(b)], pronounced differences arise: 
The sub-SNL parameter region is significantly reduced, while, for given $M_1$, the phase sensitivity no longer improves monotonically with $M_2$ -- there exists an optimal value $M_{2,\mathrm{opt}}<M_1$ that minimizes $\Delta \phi$. 
Indeed, when internal loss is present, fewer photons produced by the first TMS are available for the second TMS, resulting in a lower $M_2$ required for achieving optimal sensitivity.

It is worth noting that the curve of optimal sensitivity in Fig.~\ref{fig:two-mode}(b), $M_2=M_{2,\mathrm{opt}}(M_1)$ with optimal reference phase $\phi_0 = 0$, coincides with the dark-fringe ($\langle \hat{N}_{\text{out},2} \rangle =0$) condition  [SM Sec.~\ref{section: analogous}]
\begin{subequations}\label{eq:sen_opt}
\begin{align}
M_{2,\mathrm{opt}}(M_1) &= -\ln\left[1 - \eta_{12}+\eta_{12}e^{-M_1}\right],\label{eq:M2-opt}
\\
\left.\Delta \phi\right|_{\phi_0=0,M_{2,\mathrm{opt}}} 
&=\frac{\sqrt{1/\eta_{12}-1+e^{-M_1}}}{2\sqrt{\eta_\mathrm{det}(n_0 + 1)(e^{M_1}-1)}}.
\label{eq:phi-opt-general}
\end{align}
\end{subequations}
Equation~\eqref{eq:phi-opt-general} captures the essential sensitivity behavior of our hybrid implementation with asymmetrical loss (optical-arm efficiency $\eta_{12}$, mechanical-arm efficiency unity): 
While the sensitivity includes the detection efficiency $\eta_{\mathrm{det}}$ as just a prefactor, 
imperfect internal efficiency $\eta_{12}<1$ causes Heisenberg scaling $\propto 1/N_\mathrm{pr}$ to prevail only for $1/\eta_{12}-1\ll e^{-M_1}$, 
whereas the asymptotic scaling $\Delta \phi|_{M_1 \rightarrow \infty} = \sqrt{ (1/\eta_{\mathrm{det}})\left( 1/\eta_{12} - 1 \right)}\Delta \phi_{\mathrm{SNL}}$ is proportional to the SNL. 
Equation~\eqref{eq:phi-opt-general} presents a lower bound for the achievable phase sensitivity and implies the necessary condition $\eta_{12}>1/(1+\eta_\mathrm{det})$ for sub-SNL sensitivity, emphasizing the crucial role of maximizing $\eta_{12}$. 
Note that since the optimal condition Eq. \ref{eq:M2-opt} results in zero output intensity, the phase operating point for measuring sensitivity should be chosen in a small vicinity around the optimal value $\phi_0=0$.  

A thermally seeded mechanical mode $n_0>0$ improves the absolute sensitivity~\eqref{eq:phi-opt-general}, although more weakly than the SNL benchmark, resulting in the normalized sensitivity as in the blue and magenta curves in Fig.~\ref{fig:two-mode}(b), inset. 
Although larger values of $n_0$ impose stricter requirements for experimental control of $M_1$ and $M_2$ [SM Sec.~\ref{app:initial_phonon}], the protocol does \emph{not} require mechanical precooling to the ground state ($n_0 \ll 1$) [Fig.~\ref{fig:Sketch}(b)]. 
Crucially, however, the impact of thermal quanta entering the system \emph{during} the TMS sequence $t\in [t_1,t_4]$ can be much more adverse.

\emph{Multimode treatment including thermal decoherence}---Indeed, outside the regime of negligible thermal decoherence $n_\mathrm{th}\Gamma(t_4-t_1)\ll 1$ considered above, we must account for the 
injection of thermal quanta into the mechanical mode at rate $n_\mathrm{th}\Gamma$, which can be parametrized relative to the interaction rate with light by the quantum cooperativity $C_\mathrm{q}\equiv \mu_1/(n_\mathrm{th}\Gamma)$. 
Thermal quanta do not only map into $\hat{B}_\mathrm{out}$ and, in turn, the particular temporal mode $\hat{A}_\text{out}$ entering Eqs.~\eqref{eq:A-out_def}, but also into all the modes $\hat{A}^{(j)}_{\mathrm{out}\perp}$ orthogonal to $\hat{A}_\text{out}$ in a correlated manner.  
We tackle this multimode problem by working directly with the continuous fields $\hat{a}_{x}(t)$ and $\hat{b}_{x}(t)$ [i.e., essentially employing a time-bin mode basis]. The optical output field $\hat{a}_\text{out}(t)$, $t\in[t_3,t_4]$, resulting from the two subsequent TMS interactions, can be written in terms of the various input fields as the generic scattering relation
\begin{equation}\label{eq:a-out-multimode}
\hat{a}_{\text{out}}(t)=T_{b^{\dagger}(t_{1})}(t)\hat{b}^{\dagger}(t_{1})+\sum_j\int_{t_{1}}^{t}dt'\Phi_j(t,t')\hat{d}_{\text{in},j}(t'),
\end{equation}
where the continuum operators are $\hat{d}_{\text{in}}(t')\in\{\hat{b}_{\text{in}}^{\dagger}(t'),\hat{a}_{\text{in}}(t'),\hat{a}_{\text{int}}(t')\}$, the last of which is the intrinsic noise [cf.\ Eq.~\eqref{eq:A-in2}], and $T_{b^{\dagger}(t_{1})}(t)$ and $\Phi_j(t,t')$ are transfer functions. Note that the thermal noise $\hat{b}^\dagger_{\text{in}}(t')$ that populates both the localized and traveling modes during the first interaction $t'\in\{t_1,t_2\}$  also participates in the second interaction, leading to interference of these noise contributions in the associated transfer function $\Phi_j(t,t')$ [an analogous remark applies to $\hat{b}^\dagger(t_1)$ and $T_{b^\dagger(t_1)}(t)$].
We capture this and other effects as we derive the full multimode scattering relation~\eqref{eq:a-out-multimode} of the double-pass interaction scenario [Fig.~\ref{fig:Sketch}]; the derivation is given in SM Sec.~\ref{app:multimode}.
In SM Secs.~\ref{app:phot-count-reservoirs} and \ref{app:sensitivity} we detail how the statistical moments that determine the sensitivity~\eqref{eq:sens-SU} can be calculated using the scattering relation~\eqref{eq:a-out-multimode}.

To assess the influence of the thermal bath on the sensitivity at the fundamental level, we apply our full model to the scenario of zero technical losses, $\eta_\mathrm{tech}=1$, and assume $\tau_\mathrm{gap}=0$ in  the pulse scheme  Fig.~\ref{fig:Sketch}(b). 
We find the optimal sensitivity to occur when on average one thermal phonon enters per interaction $n_{\mathrm{th}}\Gamma\tau\!\approx\!1\Leftrightarrow M_{1} \!\approx\! C_{\mathrm{q}}$, establishing that our scheme also can be advantageous outside the regime of negligible thermal decoherence $n_{\mathrm{th}}\Gamma(t_4-t_1)\ll 1$ provided $C_\mathrm{q}\gtrsim 5$  
(see SM Sec.~\ref{SM-sec:thermal-decoh} for details).

\emph{Experimental Feasibility}---To assess the experimental feasibility of the scheme, we evaluate a setup based on a two-dimensional optomechanical crystal cavity on the basis of the full model~\eqref{eq:a-out-multimode}.  
The set of system parameters is chosen similar to those of Ref.~\cite{ren_two-dimensional_2020}: mechanical resonance $\Omega/2\pi=10$\,GHz, damping rate $\Gamma/2\pi =  8$\,Hz ($Q\equiv\Omega/\Gamma = 10^9$) at room temperature, cavity linewidth $\kappa/2\pi= 1.19$\,GHz, and drive-enhanced optomechanical coupling rate $g(t\in[t_1,t_2])/2\pi=30$\,MHz implying the interaction rate $\mu_1/2\pi=3$\,MHz. In addition, we assume lossless input and output coupling in the optomechanical system; while this poses a technical challenge, it does not constitute a fundamental limitation. 

Although increasing the pump power is an effective way to enhance the squeezing strength $M$ without introducing additional losses, this approach faces practical limitations, e.g., emerging nonlinear effects \cite{brawley_nonlinear_2016, leijssen_nonlinear_2017}, optomechanical system heating \cite{chan_laser_2011},  and parametric instability \cite{kippenberg_cavity_2008}. Instead, we consider how the squeezing strength can be increased by extending the interaction time $\tau$ while keeping $\mu_1$ fixed. 
However, this approach increases both the (mean) number of thermal phonons $n_\mathrm{th}\Gamma(2\tau+\tau_\mathrm{gap})$ entering the system and the required length of the delay line, 
$L = c_\mathrm{f} (\tau+\tau_\mathrm{gap}) \approx c_\mathrm{f} M_1/\mu_1$ (assuming $\tau_\mathrm{gap}\sim 1/\kappa\ll\tau=M_1/\mu_1$, $c_\mathrm{f}$ is the speed of light in the fiber), thereby decreasing the internal efficiency $\eta_{12}$ via $\eta_\mathrm{tech}$ (we assume the typical attenuation parameter for commercial fibers, $0.17$\,dB/km; see SM Sec.~\ref{app:fiber_loss}). 
This causes the global optimum to occur at a finite $M_1$, striking a trade-off between higher nominal squeezing strength and greater losses/decoherence, as seen in Fig.~\ref{fig:two-mode}(c), where
the resulting SNL-normalized sensitivity is presented. 

The ability to control experimental parameters  
is important for realizing the predicted performance of the scheme. 
To assess this aspect 
we account for fluctuations in the parameters $M_1$, $M_2$, $\eta_\mathrm{tech}$, and $\phi_0$ in SM Sec.~\ref{app:error_propagation}, finding that 
a 0.1\% relative error in each can result in a sensitivity 
degradation of 5\%, 34\%, and 127\% for $M_1=M_2=M=1$, 2, and 3, respectively, pointing to a potential bottleneck at higher squeezing strengths.

In conclusion, we have proposed an optomechanical SU(1,1) interferometer, thoroughly investigating its performance with regard to imperfections characteristic of such a hybrid implementation: 
internal mode mismatch, asymmetric loss, and thermal decoherence. 
By deriving an analytical expression for the optimal shaping of the drive pulses, we have demonstrated that internal mode mismatch—a primary bottleneck in multimode interferometry—can be eliminated. 
The impact of highly asymmetric internal loss in combination with vastly asymmetric bath occupancies (optical vacuum $n_\mathrm{opt}\approx 0$ versus mechanical thermal $n_\mathrm{th}\gg 1$) was quantified.  
Our analysis reveals that, even under realistic conditions such as room-temperature operation, our proposed interferometric scheme can achieve sensitivities surpassing the shot-noise limit.

The presented scheme and its comprehensive modeling, while specifically analyzed in an optomechanical context, offer a general methodology applicable to other hybrid quantum systems, such as optically-probed atom-ensembles, significantly expanding the practical applicability of multimode SU(1,1) interferometers. Our work thus lays a robust theoretical foundation for future experimental realizations, paving the way toward quantum-enhanced sensing technologies and deeper exploration of quantum measurements on localized oscillators.

\begin{acknowledgments}
We are indebted to S. A. Fedorov for his insights regarding the temporal mode structure of optomechanical interactions. We also acknowledge E. S. Polzik for useful discussions. P.R.S. acknowledges the financial support of the Deutsche Forschungsgemeinschaft (DFG) via Project SH 1228/3-1 and via the TRR 142/3 (Project No.\ 231447078, Subproject No.\ C10). C.M. acknowledges funding from the European Union's Horizon Europe research and innovation programme under the Marie Sk\l{}odowska-Curie grant agreement No.\ 101110196. 
This work is supported by VILLUM FONDEN under a Villum Investigator Grant, grant no.\ 25880, and by the Novo Nordisk Foundation through Copenhagen Center for Biomedical Quantum Sensing, grant number NNF24SA0088433. 
\end{acknowledgments}

\bibliography{My_Library3}

\appendix
\clearpage
\newpage

\setcounter{equation}{0}
\setcounter{figure}{0}
\renewcommand{\theequation}{S\arabic{equation}}
\renewcommand{\thefigure}{S\arabic{figure}}

\onecolumngrid

\section*{Supplementary Material}

\subsection{Pulsed optomechanics}\label{app:shaping}

\subsubsection{Optomechanical equation of motion and input-ouput relation}\label{app:OM-EOM-IO}

To derive the mode shaping, we consider the interaction picture and neglect the off-resonant terms by using
\begin{equation}
  \hat{a} \rightarrow \hat{a}e^{i (\Omega_L - \Omega) t}, \ \ \hat{b} \rightarrow \hat{b}e^{i \Omega t},
\end{equation}
where $\Omega_{L}$ is the laser frequency and $\Omega$ is the mechanical resonance frequency.
In this picture, we obtain the two-mode squeezing (TMS) Hamiltonian
\begin{equation}
	\hat{H}=\hbar  g(t) \left(\hat{a}^{\dagger }\hat{b}^{\dagger }+\hat{a} \hat{b}\right).
\end{equation}
The time-dependent, drive-enhanced optomechanical interaction strength $g(t)\propto\alpha(t)$ is used for mode shaping by means of the coherent intracavity drive amplitude $\alpha(t)$; the (zero-mean) operators $\hat{a}$ and $\hat{a}^\dagger$ represent the quantum fluctuations of the field. Using the Langevin equation, the equations of motion for the system are derived [see Eq.~\eqref{eq:eq_motion} in the main text]. In the regime of weak optomechanical coupling, i.e., $g \ll \kappa$, the cavity dynamics can be adiabatically eliminated. This yields
\begin{equation}
\label{eq:eliminate_cavity}
	\hat{a}(t)=-i \frac{2g(t)}{\kappa }\hat{b}^{\dagger}(t)+2\sqrt{\frac{1}{\kappa }}\hat{a}_{\text{in}}(t).
\end{equation}
Substituting Eq.~(\ref{eq:eliminate_cavity}) into Eq.~(\ref{eq:eq_motion}) in the main text and assuming that the measurement rate $\mu(t)$ is much larger than the mechanical dissipation $\Gamma$ (i.e., $\mu(t) \gg \Gamma$), the mechanical evolution is governed by
\begin{equation}
\dot{\hat{b}}^{\dagger}(t)=\frac{\mu(t)}{2}\hat{b}^{\dagger}(t)+i\sqrt{\mu(t)}\hat{a}_{\text{in}}(t)+\sqrt{\Gamma}\hat{b}_{\text{in}}^{\dagger}(t).
\end{equation}
Integrating this differential equation from an initial time $t_{\text{i}}$ gives
\begin{equation}
\hat{b}^{\dagger}(t)=e^{\frac{1}{2}M(t,t_{\text{i}})}\hat{b}^{\dagger}(t_{\text{i}})+\int_{t_{\text{i}}}^{t}dt'e^{\frac{1}{2}M(t,t')}\left(i\sqrt{\mu(t')}\hat{a}_{\text{in}}(t')+\sqrt{\Gamma}\hat{b}_{\text{in}}^{\dagger}(t')\right),\label{eq:mech-evol-sol}
\end{equation}
where 
\begin{align}
M(t,t') & \equiv\int_{t'}^{t}dt''\mu(t'')\label{eq:M-def},\\
\frac{dM(t,t')}{dt} & =\mu(t).
\end{align}

The continuous output light field carries information about the mechanical state. By substituting Eq.~\eqref{eq:eliminate_cavity} into the standard input-output relation $\hat{a}_{\text{out}}(t) =-\hat{a}_{\text{in}}(t)+i\sqrt{\mu(t)}\hat{b}^{\dagger}(t)$ (and noting that $\int_{t_{\text{i}}}^{t}dt'\delta(t-t')=1/2$), we obtain
\begin{gather}
\hat{a}_{\text{out}}(t)  =-\hat{a}_{\text{in}}(t)+i\sqrt{\mu(t)}\left[e^{\frac{1}{2}M(t,t_{\text{i}})}\hat{b}^{\dagger}(t_{\text{i}})+\int_{t_{\text{i}}}^{t}dt'e^{\frac{1}{2}M(t,t')}\left(i\sqrt{\mu(t')}\hat{a}_{\text{in}}(t')+\sqrt{\Gamma}\hat{b}_{\text{in}}^{\dagger}(t')\right)\right]\nonumber \\
  =ih_{\text{out}}(t)\sqrt{e^{M(t_{\text{f}},t_{\text{i}})}-1}\hat{b}^{\dagger}(t_{\text{i}})-\int_{t_{\text{i}}}^{t}dt'\left[2\delta(t-t')+\sqrt{\mu(t)}e^{\frac{1}{2}M(t,t')}\sqrt{\mu(t')}\right]\hat{a}_{\text{in}}(t')+i\sqrt{\mu(t)}\int_{t_{\text{i}}}^{t}dt'e^{\frac{1}{2}M(t,t')}\sqrt{\Gamma}\hat{b}_{\text{in}}^{\dagger}(t'),\label{eq:a-out-t}
\end{gather}
where we have used the solution for the mechanical evolution, Eq.~(\ref{eq:mech-evol-sol}), to express the output field in terms of the input fields and the mechanical initial condition; moreover, in the last expression, we have introduced the square-normalized optical output mode which participated in the two-mode-squeezing interaction with the mechanical mode, 
\begin{equation}
h_{\text{out}}(t)\equiv\frac{\sqrt{\mu(t)}e^{\frac{1}{2}M(t,t_{\text{i}})}}{\sqrt{e^{M(t_{\text{f}},t_{\text{i}})}-1}},\label{eq:h-out-j}
\end{equation}
referenced to a given interaction interval $t\in[t_{\text{i}},t_{\text{f}}]$.

Multiplying the output field~(\ref{eq:a-out-t}) by the output mode~(\ref{eq:h-out-j}), and integrating over the interaction interval, we get
\begin{align}
\int_{t_{\text{i}}}^{t_{\text{f}}}dt\,h_{\text{out}}(t)\hat{a}_{\text{out}}(t) & =i\sqrt{e^{M(t_{\text{f}},t_{\text{i}})}-1}\hat{b}^{\dagger}(t_{\text{i}})-\int_{t_{\text{i}}}^{t_{\text{f}}}dt''\left[h_{\text{out}}(t'')+\sqrt{\mu(t'')}\int_{t''}^{t_{\text{f}}}dth_{\text{out}}(t)\sqrt{\mu(t)}e^{\frac{1}{2}M(t,t'')}\right]\hat{a}_{\text{in}}(t'')\label{eq:h-j_inout}\\
 & \quad+i\sqrt{\Gamma}\int_{t_{\text{i}}}^{t_{\text{f}}}dt''\int_{t''}^{t_{\text{f}}}dt\,h_{\text{out}}(t)\sqrt{\mu(t)}e^{\frac{1}{2}M(t,t'')}\hat{b}_{\text{in}}^{\dagger}(t'')\nonumber\\
 & =i\sqrt{e^{M(t_{\text{f}},t_{\text{i}})}-1}\hat{b}^{\dagger}(t_{\text{i}})-e^{M(t_{\text{f}},t_{\text{i}})/2}\int_{t_{\text{i}}}^{t_{\text{f}}}dt''h_{\text{in}}(t'')\hat{a}_{\text{in}}(t'')-i\sqrt{\Gamma}\int_{t_{\text{i}}}^{t_{\text{f}}}dt''\frac{2\sinh\left(\frac{1}{2}M(t'',t_{\text{f}})\right)}{\sqrt{1-e^{-M(t_{\text{f}},t_{\text{i}})}}}\hat{b}_{\text{in}}^{\dagger}(t''), 
\end{align}
where we have used that 
\begin{align}
\int_{t''}^{t_{\text{f}}}dth_{\text{out}}(t)\sqrt{\mu(t)}e^{\frac{1}{2}M(t,t'')} & =\frac{e^{-\frac{1}{2}M(t'',t_{\text{i}})}}{\sqrt{e^{M(t_{\text{f}},t_{\text{i}})}-1}}\int_{t''}^{t_{\text{f}}}dt\mu(t)e^{M(t,t_{\text{i}})}\nonumber\\
 & =\frac{e^{-\frac{1}{2}M(t'',t_{\text{i}})}}{\sqrt{e^{M(t_{\text{f}},t_{\text{i}})}-1}}\int_{t''}^{t_{\text{f}}}dt\frac{dM(t,t_{\text{i}})}{dt}e^{M(t,t_{\text{i}})}\nonumber\\
 & =\frac{e^{-\frac{1}{2}M(t'',t_{\text{i}})}}{\sqrt{e^{M(t_{\text{f}},t_{\text{i}})}-1}}\int_{M(t'',t_{\text{i}})}^{M(t_{\text{f}},t_{\text{i}})}dMe^{M}\nonumber\\
 & =\frac{1}{\sqrt{e^{M(t_{\text{f}},t_{\text{i}})}-1}}[e^{-\frac{1}{2}M(t'',t_{\text{i}})}e^{M(t_{\text{f}},t_{\text{i}})}-e^{\frac{1}{2}M(t'',t_{\text{i}})}]=-\frac{2\sinh\left(\frac{1}{2}M(t'',t_{\text{f}})\right)}{\sqrt{1-e^{-M(t_{\text{f}},t_{\text{i}})}}},
\end{align}
and reexpressed the unnormalized input light mode appearing in Eq.~(\ref{eq:h-j_inout})
\begin{align}
\tilde{h}_{\text{in}}(t'') & \equiv h_{\text{out}}(t'')+\sqrt{\mu(t'')}\int_{t''}^{t_{\text{f}}}dth_{\text{out}}(t)\sqrt{\mu(t)}e^{\frac{1}{2}M(t,t'')}\nonumber \\
 & =\frac{\sqrt{\mu(t'')}e^{\frac{1}{2}M(t'',t_{\text{i}})}}{\sqrt{e^{M(t_{\text{f}},t_{\text{i}})}-1}}+\frac{\sqrt{\mu(t'')}}{\sqrt{e^{M(t_{\text{f}},t_{\text{i}})}-1}}[e^{-\frac{1}{2}M(t'',t_{\text{i}})}e^{M(t_{\text{f}},t_{\text{i}})}-e^{\frac{1}{2}M(t'',t_{\text{i}})}]\nonumber \\
 & =e^{M(t_{\text{f}},t_{\text{i}})/2}h_{\text{in}}(t''),\label{eq:h-in-tilde}
\end{align}
in terms of the square-normalized input mode
\begin{equation}
h_{\text{in}}(t)\equiv\frac{\sqrt{\mu(t)}e^{-\frac{1}{2}M(t,t_{\text{i}})}}{\sqrt{1-e^{-M(t_{\text{f}},t_{\text{i}})}}}.\label{eq:h-in}
\end{equation}
More generally, for an arbitrary output light mode $h_{\mathrm{out},k}(t)$, the corresponding unnormalized input light mode is
\begin{align}
\tilde{h}_{\text{in},k}(t'') & \equiv h_{\text{out},k}(t'')+\sqrt{\mu(t'')}\int_{t''}^{t_{\text{f}}}dth_{\text{out},k}(t)\sqrt{\mu(t)}e^{\frac{1}{2}M(t,t'')}\\
& = \int_{t''}^{t_{\mathrm{f}}}dt\left[2\delta(t-t'')+\sqrt{\mu(t'')}\sqrt{\mu(t)}e^{\frac{1}{2}M(t,t'')} \right]h_{\mathrm{out},k}(t)\nonumber\\
& = \int_{t''}^{t_{\mathrm{f}}}dt\left[2\delta(t-t'')+2\sinh\left(M(t_{\text{f}},t_{\text{i}})/2\right)h_{\text{in}}(t'') h_{\text{out}}(t) \right]h_{\mathrm{out},k}(t).\nonumber
\end{align}
Note that any output light mode $h_{\mathrm{out}\perp}(t)$ orthogonal to $h_{\mathrm{out}}(t)$ has no contribution from $\hat{b}^\dagger(t_\mathrm{i})$ and hence does not participate in the optomechanical two-mode-squeezing interaction [as follows by projecting out $h_{\mathrm{out}\perp}(t)$ analogously to the calculation in Eq.~\eqref{eq:h-j_inout}].

The particular input light mode $h_\mathrm{in}(t)$ is of course only one out of the countably infinite orthonormal set of modes $\{h_{\mathrm{in},k}(t)\}$ required to specify the state of the input light field $\hat{a}_\mathrm{in}(t)=\sum_k \hat{a}_{\mathrm{in},k} h_{\mathrm{in},k}(t)$ on the interval $t\in [t_\mathrm{i},t_\mathrm{f}]$.
An arbitrary input light mode $h_{\mathrm{in},k}(t)$ maps, according to the transformation in Eq.~\eqref{eq:a-out-t}, to the unnormalized output light mode 
\begin{align}
\tilde{h}_{\mathrm{out},k}(t) & =\int_{t_{\text{i}}}^{t}dt'\left[2\delta(t-t')+\sqrt{\mu(t)}e^{\frac{1}{2}M(t,t')}\sqrt{\mu(t')}\right]h_{\mathrm{in},k}(t')\nonumber \\
 & =\int_{t_{\text{i}}}^{t}dt'\left[2\delta(t-t')+2\sinh\left(M(t_{\text{f}},t_{\text{i}})/2\right)h_{\text{out}}(t)h_{\text{in}}(t')\right]h_{\mathrm{in},k}(t'),\label{eq:light-transform-h-in-out}
\end{align}
in terms of the normalized input and output modes involved in the two-mode-squeezing process, Eqs.~\eqref{eq:h-out-j} and \eqref{eq:h-in}. Applying this to the particular input light mode~\eqref{eq:h-in} identified above, $h_{\mathrm{in},k}(t)=h_{\mathrm{in}}(t)$, 
 \begin{align}
 \tilde{h}_{\mathrm{out}}(t) & =\frac{1}{\sqrt{1-e^{-M(t_{\text{f}},t_{\text{i}})}}}\left(\sqrt{\mu(t)}e^{-\frac{1}{2}M(t,t_{\text{i}})}+\sqrt{\mu(t)}\int_{t_{\text{i}}}^{t}dt'\sqrt{\mu(t')}e^{-\frac{1}{2}M(t',t_{\text{i}})}\sqrt{\mu(t')}e^{\frac{1}{2}M(t,t')}\right)\nonumber \\
 & =\frac{\sqrt{\mu(t)}}{\sqrt{1-e^{-M(t_{\text{f}},t_{\text{i}})}}}\left(e^{-\frac{1}{2}M(t,t_{\text{i}})}+e^{\frac{1}{2}M(t,t_{\text{i}})}\int_{t_{\text{i}}}^{t}dt'\mu(t')e^{-M(t',t_{\text{i}})}\right)\nonumber \\
 & =\frac{\sqrt{\mu(t)}}{\sqrt{1-e^{-M(t_{\text{f}},t_{\text{i}})}}}\left(e^{-\frac{1}{2}M(t,t_{\text{i}})}+e^{\frac{1}{2}M(t,t_{\text{i}})}\int_{0}^{M(t,t_{\text{i}})}dMe^{-M}\right)\nonumber \\
 & =e^{M(t_{\text{f}},t_{\text{i}})/2}h_{\text{out}}(t),\label{eq:h-in-to-out}
\end{align}
showing that the temporal input mode $h_{\text{in}}(t)$ maps exclusively to the output mode $h_{\text{out}}(t)$ [this can also be shown by exploiting the unitarity of Eq.~(\ref{eq:a-out-t}) in the limit $\Gamma\rightarrow0$]. The above observations establish that, under circumstances where the mechanical thermal noise contribution $\propto\hat{b}_\mathrm{in}^\dagger$ in Eq.~\eqref{eq:a-out-t} is negligible, the pulsed interaction of our optomechanical system is fully captured by the two-mode description employed in the main text, with the input and output operators of a single TMS interaction pulse given by
\begin{equation}
\hat{A}_{\text{out(in)}}=\int_{t_\mathrm{i}}^{t_\mathrm{f}} \mathrm{d}t h_\mathrm{out(in)}(t) \hat{a}_{\text{out(in)}}(t), \label{eq-app:A-out-in}
\end{equation}
in terms of the mode envelopes~\eqref{eq:h-out-j} and \eqref{eq:h-in} identified above.

Using Eq.~(\ref{eq:h-in}) we can write the integral of the square-normalized input mode over a subinterval of $[t_\mathrm{i},t_\mathrm{f}]$ as
\begin{equation}
\int_{t_{\text{i}}}^{t}dt'h_{\text{in}}^{2}(t') =\frac{\int_{t_{\text{i}}}^{t}dt'\mu(t')e^{-M(t',t_{\text{i}})}}{1-e^{-M(t_{\text{f}},t_{\text{i}})}}=\frac{1-e^{-M(t,t_{\text{i}})}}{1-e^{-M(t_{\text{f}},t_{\text{i}})}}.
\end{equation}
From this we can express $M(t,t_{\text{i}})$ and its derivative $\mu(t)=dM(t,t_{\text{i}})/dt$ in order to find $\mu(t)$ in terms of a prescribed, square-normalized input mode envelope $h_{\text{in}}(t)$ 
\begin{align}
M(t,t_{\mathrm{i}}) & =-\ln\left(1-(1-e^{-M(t_{\mathrm{f}},t_{\mathrm{i}})})\int_{t_{\mathrm{i}}}^{t}dt'h_{\mathrm{in}}^{2}(t')\right)\Rightarrow\label{eq:M-from-h-in}\\
\mu(t) & =\frac{h_{\mathrm{in}}^{2}(t)}{\frac{1}{1-e^{-M(t_{\mathrm{f}},t_{\mathrm{i}})}}-\int_{t_{\mathrm{i}}}^{t}dt'h_{\mathrm{in}}^{2}(t')}.\label{eq:mu-from-h-in}
\end{align}
Analogously, we can (for reference) derive expressions for the $M(t,t_\mathrm{i})$ and $\mu(t)$ required to realize a prescribed, square-normalized output mode envelope $h_{\text{out}}(t)$; observing from Eq.~\eqref{eq:h-out-j} that
\begin{equation}
\int_{t_{\mathrm{i}}}^{t}dt'h_{\mathrm{out}}^{2}(t') =\frac{\int_{t_{\mathrm{i}}}^{t}dt'\mu(t')e^{M(t',t_{\mathrm{i}})}}{e^{M(t_{\mathrm{f}},t_{\mathrm{i}})}-1}=\frac{e^{M(t,t_{\mathrm{i}})}-1}{e^{M(t_{\mathrm{f}},t_{\mathrm{i}})}-1},
\end{equation}
we find
\begin{align}
M(t,t_{\mathrm{i}})	&=\ln\left(1+(e^{M(t_{\mathrm{f}},t_{\mathrm{i}})}-1)\int_{t_{\mathrm{i}}}^{t}dt'h_{\mathrm{out}}^{2}(t')\right)\Rightarrow\\
\mu(t) &=\frac{h_{\mathrm{out}}^{2}(t)}{\frac{1}{e^{M(t_{\mathrm{f}},t_{\mathrm{i}})}-1}+\int_{t_{\mathrm{i}}}^{t}dt'h_{\mathrm{out}}^{2}(t')}.
\end{align}

\subsubsection{Shaping of optical input and output modes for two concatenated interactions}\label{app:shaping-sub}

We consider two consecutive optomechanical two-mode-squeezing interactions according to the protocol sketched in main-text Fig.~\ref{fig:Sketch}. These are of equal duration $t_2-t_1 = t_4-t_3 \equiv \tau$ and their respective optical input and output modes are engineered via individual drive envelopes $\mu_1(t)$ and $\mu_2(t)$ for $t\in [0,\tau]$, leading to the drive envelope on the full interval of the protocol $t\in [t_1,t_4]$
\begin{equation}\label{eq-app:mu-global}
    \mu(t) = 
    \begin{cases}
        \mu_1(t-t_1) & t\in [t_1,t_2] \\
        0 & t\in [t_2,t_3] \\
        \mu_2(t-t_3) & t\in [t_3,t_4]
    \end{cases}.
\end{equation}
Accordingly, we can introduce the squeezing strength functions $M_j(t)$ with $t\in[0,\tau]$ of the individual interactions 
\begin{align}
M_1(t) &\equiv M(t+t_1,t_1)=\int_0^t dt' \mu_1(t') \\
M_2(t) &\equiv M(t+t_3,t_3)=\int_0^t dt' \mu_2(t'),
\end{align}
cf.\ Eq.~\eqref{eq:M-def}; we use the notation $M_j\equiv M_j(\tau)$ for the net strength of interaction $j$.
We employ the following notation for the input and output operators of the first and second interaction, respectively,
\begin{align}
\hat{A}^{(1)}_{\text{out(in)}}&=\int_{t_1}^{t_2} \mathrm{d}t h^{(1)}_\mathrm{out(in)}(t-t_1) \hat{a}_{\text{out(in)}}(t)\\
\hat{A}^{(2)}_{\text{out(in)}}&=\int_{t_3}^{t_4} \mathrm{d}t h^{(2)}_\mathrm{out(in)}(t-t_3) \hat{a}_{\text{out(in)}}(t),
\label{eq-app:A-out-in-j}
\end{align}
where the temporal mode functions $h^{(j)}_\mathrm{out(in)}(t)$ ($t\in[0,\tau]$) are given by 
\begin{subequations}\label{eq-app:h-out-in-j}
\begin{align}
h^{(j)}_{\text{out}}(t)&\equiv\frac{\sqrt{\mu_j(t)}e^{\frac{1}{2}M_j(t)}}{\sqrt{e^{M_j}-1}}\label{eq-app:h-out-j}\\
h^{(j)}_{\text{in}}(t)&\equiv\frac{\sqrt{\mu_j(t)}e^{-\frac{1}{2}M_j(t)}}{\sqrt{1-e^{-M_j}}},\label{eq-app:h-in-j}
\end{align}
\end{subequations}
cf.\ Eqs.~\eqref{eq:h-out-j} and \eqref{eq:h-in}.

We now address the matching of the optical input mode of the second interaction to the output mode of the first $h^{(2)}_\mathrm{in}(t)=h^{(1)}_\mathrm{out}(t)$. Assuming a given choice of drive envelope for the first interaction $\mu_1(t)$, the required choice for $\mu_2(t)$ follows from inserting $h_\mathrm{in}(t)=h^{(1)}_\mathrm{out}(t-t_3)$ [Eq.~\eqref{eq-app:h-out-j}] into Eq.~\eqref{eq:mu-from-h-in} with $t_\mathrm{i}=t_3$ and $t_\mathrm{f}=t_4$, 
\begin{equation}
\mu_{2}(t) = \frac{\mu_{1}(t)}{\frac{e^{M_{1}}-e^{-M_{2}}}{1-e^{-M_{2}}}e^{-\int_{0}^{t}dt'\mu_{1}(t')}-1},\label{eq-app:mu2-sol-mu1}
\end{equation}
in view of Eq.~\eqref{eq-app:mu-global}. Equation~\eqref{eq-app:mu2-sol-mu1} establishes the family of drive-pulse pairs $\{\mu_1(t),\mu_2(t)\}$ that lead to internal mode matching $h^{(2)}_\mathrm{in}(t)=h^{(1)}_\mathrm{out}(t)$.

Considering the particular case where the drive envelope over the duration of the first interaction $\mu_1(t)=\mu_1$ is a constant, Eq.~\eqref{eq-app:mu2-sol-mu1} shows that we must shape the pulse as [Eq.~\eqref{eq:mu2-shape} in the main text]
%
\begin{equation}
\label{eq:mu2}
\mu_2(t)=\frac{\mu_1}{\frac{e^{M_1}-e^{-M_2}}{1-e^{-M_2}}e^{-\mu_1 t}-1}\, ,
\end{equation}
the peak value of which occurs at the end of the drive pulse 
\begin{equation}
    \mu_2(\tau) = \mu_1 e^{M_2}\frac{1-e^{-M_2}}{1-e^{-M_1}}\, .
\end{equation}
The mode shapes~\eqref{eq-app:h-out-in-j} resulting from this pair of drive envelopes is sketched in main-text Fig.~\ref{fig:Sketch}.

\subsection{Performance in the absence of mode shaping 
\label{app:without_shaping}}

The matching of the input mode of the second interaction and the output mode of the first interaction, described in SM Sec.~\ref{app:shaping-sub} is essential for achieving appreciable sub-SNL performance using the SU(1,1) scheme. To see this, we consider the mode overlap coefficient which describe mismatch losses
\begin{equation}\label{eq-SM:eta-mode}
    \eta_{\mathrm{mode}} = \int_{0}^{\tau} h_{\mathrm{out}}^{(1)}(t)h_{\mathrm{in}}^{(2)}(t) d t.
\end{equation}

\begin{figure}[H]
\centering
\includegraphics[width=0.8\columnwidth]{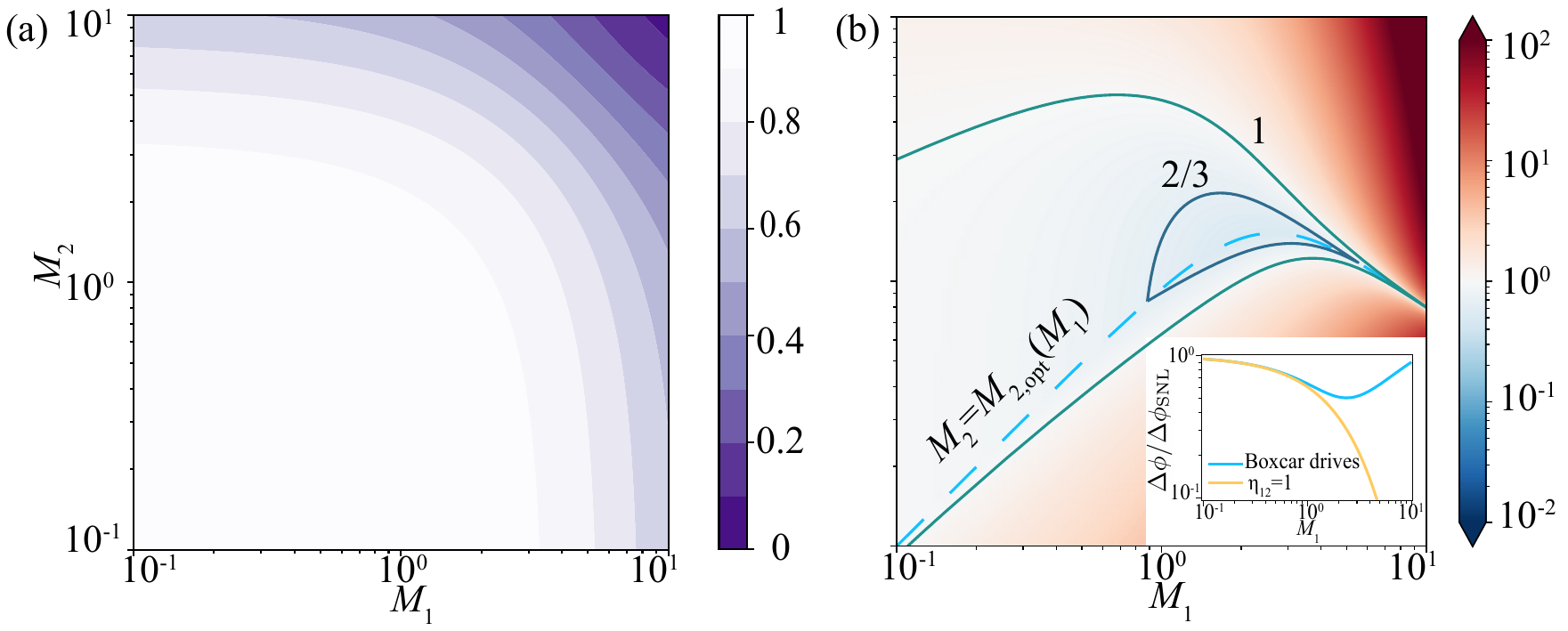}
\caption{\label{fig:sen_boxcar} Characterization of an SU(1,1) interferometer without mode matching. (a) Mode overlap coefficient $\eta_{\mathrm{mode}}$ as a function of $M_1$ and $M_2$ for dual boxcar driving according to Eq.~(\ref{eq:eta12-boxcars}). (b) Phase sensitivity optimized over $\phi_0$ for a pair of boxcar drive envelopes, where $\eta_{\mathrm{mode}}$ is taken as in Eq.~\eqref{eq:eta12-boxcars}. The dashed line marks the optimal strength of the second interaction, $M_{2,\mathrm{opt}}(M_1)$, which reflects the dependence $\eta_{\mathrm{mode}}(M_1,M_2)$. The inset compares the phase sensitivity at this optimum with that obtained under perfect mode matching.}
\end{figure}
As a benchmark for the importance of mode shaping, we consider the naive choice of a pair of boxcar drive-pulse envelopes, i.e., where $\mu_1(t)=\mu_1$ and $\mu_2(t)=\mu_2$ are constants on the interval $t\in[0,\tau]$ and zero elsewhere. In this case, the mode shapes~\eqref{eq-app:h-out-in-j} are exponential functions, for which Eq.~\eqref{eq-SM:eta-mode} results in
\begin{equation}\label{eq:eta12-boxcars}
    \eta_{\mathrm{mode}} = \sqrt{\frac{M_{1}M_{2}/4}{\sinh(M_{1}/2)\sinh(M_{2}/2)}}\frac{\sinh([M_{1}-M_{2}]/4)}{[M_{1}-M_{2}]/4} \xrightarrow{M_1=M_2\equiv M} \frac{M/2}{\sinh(M/2)},
\end{equation}
which is plotted in Fig.~\ref{fig:sen_boxcar}(a).
One can observe that in the low-strength regime $\eta_{\mathrm{mode}} \rightarrow 1$. However, surpassing the SNL in this regime can only be minor as shown in Fig.~\ref{fig:ideal_case}(a) in the main text. In the high-strength regime, $\eta_{\mathrm{mode}}$ decreases quickly, and whenever $\eta_{\mathrm{mode}}<1/2$, there is no possibility to surpass SNL, as can be seen in Eq.~(\ref{eq:phi-opt-general}) in the limit $M_1 \rightarrow \infty$ and assuming $n_0=0$ and $\eta_{\mathrm{det}}=1$. For equal strengths $M_1=M_2\equiv M$, the efficiency~\eqref{eq:eta12-boxcars} crosses the threshold $\eta_{\mathrm{mode}}<1/2$ for $M\gtrapprox 4.4$. Hence, only by employing proper mode shaping one can substantially overcome the SNL under realistic experimental conditions.

The effect of the boxcar driving envelopes on the phase sensitivity is illustrated in Fig.~\ref{fig:sen_boxcar}(b). In the low-strength regime, the sensitivity closely resembles that of the idealized case shown in Fig.~\ref{fig:ideal_case}(a), because the mode-matching coefficient $\eta_\mathrm{mode}$ remains close to unity. However, as the squeezing strength increases, the mode matching coefficient decreases, leading to an increase in internal losses; therefore, the phase sensitivity behaves similarly to Fig.~\ref{fig:ideal_case}(b).

\subsection{Matrix formalism for idealized two-mode-squeezing interaction}
\label{section: analogous}

In the main text, we present first the analysis of the SU(1,1) interferometer in the regime where the two nonlinear optomechanical interaction processes are not competing with simultaneous mechanical thermal decoherence, as described by Eq.~(\ref{eq:2-mode-IO}). However, we do take into account both internal and external optical losses as well as the initial thermal occupancy of the mechanical mode. Notably, in our analysis, mode mismatch characterized by Eqs.~(\ref{eq:A-in2}) is incorporated as a source of internal loss. 
Here, we introduce a matrix formalism that encapsulates the input-output relations, the phase shift and aforementioned losses, and allows to explicitly calculate the phase sensitivity in this few-mode regime. 

The input-output relations for the temporal modes of the two-mode model described in Eq.~(\ref{eq:2-mode-IO}) can be expressed in a compact matrix form as:
\begin{equation}  \mathbf{A}_{\text{out}}= \mathbf{P} \mathbf{A}_{\text{in}},
\end{equation}
where
\begin{equation}
    \begin{aligned}
        \mathbf{A}_{\text{in}(\text{out})}&=\left(
\begin{array}{c}
 A_{\text{in}(\text{out})} \\
 B_{\text{in}(\text{out})}^{\dagger } \\
\end{array}
\right)
    \end{aligned},
\end{equation}
\begin{equation}
    \begin{aligned}
       \pmb{\text{P}}\pmb{=}\left(
\begin{array}{cc}
 -\sqrt{e^M} & i \sqrt{e^M-1} \\
 i \sqrt{e^M-1} & \sqrt{e^M} \\
\end{array}
\right).
    \end{aligned}
\end{equation}
Here, $M$ is the interaction strength parameter of the two-mode-squeezing (TMS) process. 

By considering two sequential TMS interactions, we can express the overall input-output relation for the two stages using subscripts to distinguish the stages. The relation takes the form:
\begin{equation}
\mathbf{A}_{\text{out},2}\pmb{=}\mathbf{{P}}_2\mathbf{S}\mathbf{{P}}_1\mathbf{A}_{\text{in},1} = \begin{pmatrix}
\sqrt{e^{M_{1}+M_{2}}} - e^{i\phi}\sqrt{e^{M_{1}}-1}\sqrt{e^{M_{2}}-1} &
-i(\sqrt{e^{M_{2}}}\sqrt{e^{M_{1}}-1} - e^{i\phi}\sqrt{e^{M_{1}}}\sqrt{e^{M_{2}}-1})\\[6pt]
i(e^{i\phi}\sqrt{e^{M_{2}}}\sqrt{e^{M_{1}}-1} - \sqrt{e^{M_{1}}}\sqrt{e^{M_{2}}-1}) &
\sqrt{e^{M_{1}+M_{2}}}e^{i\phi} - \sqrt{e^{M_{1}}-1}\sqrt{e^{M_{2}}-1}
\end{pmatrix}
\mathbf{A}_{\text{in},1},
\end{equation}
where $\mathbf{S}$ represents a phase shift applied to mode \(B\) and is given by
\(\mathbf{S}=\left(
\begin{array}{cc}
 1 & 0 \\
 0 & e^{i \phi } \\
\end{array}
\right)\).

To account for losses, we introduce two types of efficiencies:	1)	internal efficiency of the optical arms between the two interactions $\bm{\eta}_{12}$ and 2) detection efficiency at the output of the protocol $\bm{\eta}_{\text{det}}$, which is associated with external detection losses. Considering such losses, the  modified input-output relation, including the contributions from vacuum noise, becomes
\begin{equation}\label{eq:A-out2-vec}
\mathbf{A}_{\text{out},2}\pmb{=}\sqrt{{\bm{\eta}}_{\text{det}}}\mathbf{{P}}_2 (\sqrt{\bm{\eta}_{12}} \mathbf{{S}}\mathbf{{P}}_1\mathbf{A}_{\text{in},1}+\sqrt{\mathbf{I}-\bm{\eta}_{12}}\mathbf{A}_{\text{vac,1}})+\sqrt{\mathbf{I}-\bm{\eta}_{\text{det}}}\mathbf{A}_{\text{vac,2}},
\end{equation}
where
\begin{equation}\label{eq:A-vac1-vec}
    \begin{aligned}
        \mathbf{A}_{\text{vac,1}(\text{vac,2})}&=\left(
\begin{array}{c}
 \hat{A}_{\text{vac,1}(\text{vac,2})} \\
 0 \\
\end{array}
\right)
    \end{aligned},
\end{equation}
represents the vacuum noise contributions. The internal and detection efficiency matrices read 
\begin{equation}
\bm{\eta}_{12}\pmb{=}\left(
\begin{array}{cc}
 {\eta_{12}} & 0 \\
 0 & {1 } \\
\end{array}
\right),\quad
\bm{\eta}_{\text{det}}\pmb{=}\left(
\begin{array}{cc}
{\eta_{\text{det}}} & 0 \\
 0 & {1} \\
\end{array}
\right).
\end{equation}
The $\mathbf{A}_{\text{in},1}$-dependent contribution in Eq.~(\ref{eq:A-out2-vec}) can be written explicitly as
\begin{multline}\label{app-eq:transferM-loss}
\sqrt{{\bm{\eta}}_{\text{det}}}\mathbf{{P}}_2 \sqrt{\bm{\eta}_{12}} \mathbf{{S}}\mathbf{{P}}_1\mathbf{A}_{\text{in},1}
=\\
\begin{pmatrix}
\sqrt{\eta_{\mathrm{det}}}\bigl(\sqrt{\eta_{12}}\sqrt{e^{M_{1}+M_{2}}}
   - e^{i\phi}\sqrt{e^{M_{1}}-1}\,\sqrt{e^{M_{2}}-1}\bigr)
&
-\,i\sqrt{\eta_{\mathrm{det}}}\bigl(\sqrt{\eta_{12}}\sqrt{e^{M_{2}}}\,\sqrt{e^{M_{1}}-1}
   - e^{i\phi}\sqrt{e^{M_{1}}}\,\sqrt{e^{M_{2}}-1}\bigr)
\\[6pt]
i\bigl(e^{i\phi}\sqrt{e^{M_{2}}}\,\sqrt{e^{M_{1}}-1}
   - \sqrt{\eta_{12}}\sqrt{e^{M_{1}}}\,\sqrt{e^{M_{2}}-1}\bigr)
&
e^{i\phi}\sqrt{e^{M_{1}+M_{2}}}
   - \sqrt{\eta_{12}}\sqrt{e^{M_{1}}-1}\,\sqrt{e^{M_{2}}-1}
\end{pmatrix}
\mathbf{A}_{\text{in},1}.
\end{multline}
Specifically, Eq.~(\ref{eq:A-out2-vec}) expresses the optical output operator $\hat{A}_{\mathrm{out,2}}$ as a linear combination of the optical-vacuum input operators $\hat{A}_{\mathrm{in,1}}$, $\hat{A}_{\mathrm{vac,1}}$, and $\hat{A}_{\mathrm{vac,2}}$ and the thermal mechanical input operator $\hat{B}_{\mathrm{in,1}}^{\dagger}$, which are mutually uncorrelated. However, since the transfer coefficients of these input fields into the output field $\hat{A}_{\mathrm{out,2}}$ are constrained by unitarity, the transfer coefficients of the optical vacuum fields can be eliminated from the calculation, as shown in SM Sec.~\ref{app:sensitivity}. In fact, in the limit $\Gamma n_\mathrm{th}\rightarrow 0$, the performance of the scheme can, as shown in SM Sec.~\ref{app-sec:sens-noMechDecoh}, be evaluated solely on the basis of the mechanics-to-optics transfer coefficient $T_\mathrm{MO}$, $\hat{A}_{\mathrm{out},2} = T_\mathrm{MO} \hat{B}^\dagger_{\mathrm{in},1}+\ldots$, i.e., the upper-right element of the matrix seen in Eq.~\eqref{app-eq:transferM-loss} [dots indicate the remaining terms given by Eq.~\eqref{eq:A-out2-vec}], and the initial thermal mean occupancy of the mechanical mode $\langle \hat{B}_{\mathrm{in,1}}^{\dagger} \hat{B}_{\mathrm{in,1}}\rangle \equiv n_0$. 
Hence, to evaluate the sensitivity $\Delta \phi$, as given by Eq.~\eqref{eq:sens-SU} in the main text, we calculate the first moment $\langle \hat{N}_{\mathrm{out,2}} \rangle$ and its derivative 
\begin{subequations}\label{eq:expect}
\begin{gather}
    \langle \hat{N}_{\mathrm{out,2}} \rangle = \langle \hat{A}_{\mathrm{out,2}}^{\dagger} \hat{A}_{\mathrm{out,2}}\rangle = (n_0 + 1)|T_\mathrm{MO}|^2 = (n_0 +1) \eta_\mathrm{det}\left|\sqrt{e^{M_{2}}-1}e^{-i\phi}e^{M_{1}/2}-\sqrt{\eta_{12}}e^{M_{2}/2}\sqrt{e^{M_{1}}-1}\right|^{2} \Rightarrow \label{eq-app:N-out2-exp-2mode}   \\
    \frac{d\langle \hat{N}_{\mathrm{out,2}} \rangle}{d\phi} =2(n_0 +1)\eta_\mathrm{det}\sqrt{\eta_{12}}\sqrt{e^{M_{2}}-1}e^{M_{2}/2}\sqrt{e^{M_{1}}-1}e^{M_{1}/2}\sin(\phi),
\end{gather}
\end{subequations}
and the variance [Eq.~\eqref{eq-app:VarN-noMechDecoh2}]
\begin{equation}\label{eq-SM:Var-out_noDecoh}
    \mathrm{Var}[\hat{N}_{\mathrm{out,2}}] = \langle \hat{N}_{\mathrm{out,2}}^{2}  \rangle - \langle \hat{N}_{\mathrm{out,2}}  \rangle^{2} = \langle \hat{N}_{\mathrm{out,2}} \rangle(\langle \hat{N}_{\mathrm{out,2}} \rangle + 1),
\end{equation}
which is linked to the first moment~\eqref{eq-app:N-out2-exp-2mode} by the relation that corresponds to thermal states [last expression in Eq.~\eqref{eq-SM:Var-out_noDecoh}].
This results in the expression for the sensitivity 
\begin{equation}
    \Delta \phi = \frac{\sqrt{\left|\sqrt{e^{M_{2}}-1}e^{-i\phi_0}e^{M_{1}/2}-\sqrt{\eta_{12}}e^{M_{2}/2}\sqrt{e^{M_{1}}-1}\right|^{2}\left(\left|\sqrt{e^{M_{2}}-1}e^{-i\phi_0}e^{M_{1}/2}-\sqrt{\eta_{12}}e^{M_{2}/2}\sqrt{e^{M_{1}}-1}\right|^{2}+\frac{1}{\eta_\mathrm{det}(n_0 + 1)}\right)}}{2\sqrt{\eta_{12}}\sqrt{e^{M_{2}}-1}e^{M_{2}/2}\sqrt{e^{M_{1}}-1}e^{M_{1}/2}|\sin(\phi_0)|}.
\end{equation}
By optimizing over the phase $\phi_0$, we obtain the optimum sensitivity for given $(M_1,M_2)$. The explicit analytical form of the optimal $\phi_0$ is lengthy and not physically illuminating, so we do not reproduce it here; however, we use it to generate Fig.~\ref{fig:two-mode}(a,b). Furthermore, optimizing over $M_2$ for given $M_1$, we find the optimum point $M_{2,\mathrm{opt}}(M_1)$ and the corresponding optimum sensitivity given in Eqs.~\eqref{eq:sen_opt} in the main text.

As a curiosity, we remark that Eq.~(\ref{eq:M2-opt}) can be understood as an effective squeezing strength of the lossy system: When squeezed light with the quadrature variance $e^{-r}$ suffers losses due to a finite transmission $\eta$, its quadrature variance becomes $1-\eta + \eta e^{-r}$,
from which the effective squeezing strength of the lossy system can be defined as  $\left.M_{1,\mathrm{eff}} = -\ln[1 - \eta_{12} + \eta_{12} e^{-M_1}]\right.$, 
determining the optimal strength of the second interaction $M_{2,\mathrm{opt}} = M_{1,\mathrm{eff}}$. In the absence of internal loss, i.e., $\eta_{12}=1$, Eq.~\eqref{eq:M2-opt} results in  $M_{\mathrm{2,opt}} = M_1$, while  $\Delta \phi_{\mathrm{opt}}$ in Eq.~(\ref{eq:phi-opt-general}) coincides with the inset in Fig.~\ref{fig:two-mode}(a).

\subsection{Influence of the initial thermal phonon occupation $n_0$}
\label{app:initial_phonon}

\begin{figure}[h]
\includegraphics[width=0.49\columnwidth]{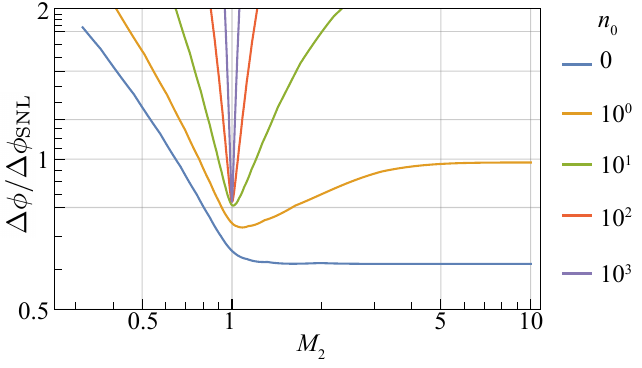}
\caption{\label{fig:n0_dependence}SNL-normalized sensitivity $\Delta \phi / \Delta \phi_\mathrm{SNL}$ in the presence of thermal decoherence for $M_1$ = 1 as a function of $M_2$ for different values of the initial phonon number $n_0$. As $n_0$ increases, the minimum sensitivity increases from approximately 0.6, and eventually saturates near 0.8, as can be qualitatively inferred from Eq.~(\ref{eq:phi-opt-general}). Additionally, the sensitivity valley broadens with decreasing $n_0$. This indicates that the required experimental precision in controlling the squeezing strength increases when the precooling is less effective. The parameters used are $n_{\mathrm{th}}=6\times 10^2$, $\Gamma/2\pi =  8$\,Hz, and an optical loss of 0.17 dB/km in the delay line, identical to those of the full-model results in Fig.~\ref{fig:ideal_case}(c), except for the variation in $n_0$. }
\end{figure}

An important practical question is to what extent mechanical precooling is necessary for effective operation of the SU(1,1) interferometer. Fig.~\ref{fig:n0_dependence} presents the results from the complete model (including thermal decoherence) showing the minimum phase sensitivity as a function of $M_2$, with fixed $M_1$. We use the full model in order to assess the effect of $n_0>0$ under the most realistic circumstances possible. As shown, increasing the initial phonon number $n_0$ not only worsens the minimum rescaled sensitivity [although the \emph{absolute} sensitivity $\Delta\phi$ improves, cf.\ Eq.~\eqref{eq:phi-opt-general}], but also reduces the range of $M_2$ for which the sensitivity surpasses the SNL. This means that, for large $n_0$, precooling does not significantly improve the minimum achievable sensitivity relative to the SNL, but it does relax the requirements on the precise control of $M_2$. In other words, effective precooling to a small value of $n_0$ broadens the parameter regime for achieving optimal sensitivity, making the system less sensitive to imperfections in the tuning of the squeezing strength.

\subsection{Multimode treatment of double-pass oscillator-field interaction\label{app:multimode}}

Here we perform a full multimode calculation of an optomechanical
SU(1,1) interferometry scheme assuming temporal mode matching of
the two optomechanical interactions involved. In the optomechanical
implementation considered here, the two ``arms'' of the interferometer
are constituted by a traveling optical field and a stationary mechanical
oscillator, respectively. These two degrees of freedom interact twice
according to a two-mode-squeezing Hamiltonian during the (disjoint)
time intervals $t\in[t_{1},t_{2}]$ and $t\in[t_{3},t_{4}]$ of equal
duration, $t_{2}-t_{1}=t_{4}-t_{3}\equiv\tau$. These two interactions
are separated by a gap $\tau_{\text{gap}}\equiv t_{3}-t_{2}\geq0$
during which a small signal phase shift is applied to the mechanical
oscillator. The purpose of the scheme is to estimate this signal phase
shift.

We now use the results of SM Sec.~\ref{app:shaping} for pulsed optomechanical
interaction to construct the net scattering relation resulting from
two subsequent optomechanical interactions with the output of the
first interaction looped back to serve as the input light for the
second interaction according to the time sequence described above [and illustrated in main-text Fig.~\ref{fig:Sketch}(b)].

Starting with the first interaction, taking place in the time interval
$t\in[t_{\text{i}}=t_{1},t_{\text{f}}=t_{2}]$, we find from Eq.~(\ref{eq:mech-evol-sol})
the mechanical creation operator at the end of the interaction
\begin{equation}
\hat{b}^{\dagger}(t_{2})=e^{\frac{1}{2}M_{1}}\hat{b}^{\dagger}(t_{1})+\int_{t_{1}}^{t_{2}}dt''e^{\frac{1}{2}M(t_{2},t'')}\left(i\sqrt{\mu(t'')}\hat{a}_{\text{in}}(t'')+\sqrt{\Gamma}\hat{b}_{\text{in}}^{\dagger}(t'')\right),\label{eq:mech-evol-sol-int1}
\end{equation}
where we have defined the net interaction strength $M_{1}\equiv M(t_{2},t_{1})$
of the first interaction. For reference, we note that the mechanical
thermal occupancy at $t_{2}$ is, when assuming a constant optomechanical
coupling rate during the first interval $\mu(t\in[t_{1},t_{2}])=\mu_{1}$,
\begin{align}
\langle\hat{b}^{\dagger}(t_{2})\hat{b}(t_{2})\rangle & =n_{0}e^{M_{1}}+\int_{t_{1}}^{t_{2}}dt''e^{M(t_{2},t'')}\left(\mu(t'')+\Gamma_{\text{th}}\right)\nonumber\\
\xrightarrow{\mu(t\in[t_{1},t_{2}])=\mu_{1}=M_{1}/\tau} & =n_{0}e^{M_{1}}+\int_{t_{1}}^{t_{2}}dt''e^{\mu_{1}(t_{2}-t'')}\left(\mu_{1}+\Gamma_{\text{th}}\right)\nonumber\\
 & =n_{0}e^{M_{1}}+[e^{M_{1}}-1]\left(1+\frac{\Gamma_{\text{th}}}{\mu_{1}}\right),
\end{align}
in terms of the initial occupancy $\langle\hat{b}^{\dagger}(t_{1})\hat{b}(t_{1})\rangle\equiv n_{0}$ and the thermal decoherence rate $\Gamma_\mathrm{th}=n_\mathrm{th}\Gamma$.
From Eqs.~(\ref{eq:a-out-t}) and (\ref{eq:light-transform-h-in-out})
we see that the continuous output light field of the first interaction
is, i.e., for $t\in[t_{1},t_{2}]$,
\begin{multline}
\hat{a}_{\text{out}}(t)  =ih_{\text{out}}^{(1)}(t)\sqrt{e^{M_{1}}-1}\hat{b}^{\dagger}(t_{1})-\int_{t_{1}}^{t}dt'\left[2\delta(t-t')+2\sinh\left(M_{1}/2\right)h_{\text{out}}^{(1)}(t)h_{\text{in}}^{(1)}(t')\right]\hat{a}_{\text{in}}(t')\\+i\sqrt{\mu(t)}\int_{t_{1}}^{t}dt'e^{\frac{1}{2}M(t,t')}\sqrt{\Gamma}\hat{b}_{\text{in}}^{\dagger}(t'),\label{eq:a-out-t-int1}
\end{multline}
in terms of the output temporal mode of the first interaction, which we for clarity define in the present section as a function of the absolute time $t\in[t_1,t_2]$,
\begin{equation}
h_{\text{out}}^{(1)}(t)=\frac{\sqrt{\mu(t)}e^{\frac{1}{2}M(t,t_{1})}}{\sqrt{e^{M_{1}}-1}},\label{eq:h-out-1}
\end{equation}
i.e., as in Eq.~(\ref{eq:h-out-j}). Indeed, we will throughout the present section use the absolute time axis $[t_1,t_4]$, shown in main-text Fig.~\ref{fig:Sketch}(b), for defining the temporal modes $h_{\mathrm{out(in)}}^{(j)}(t)$ and the drive envelope $\mu(t)$.

In the time interval $t\in[t_{2},t_{3}]$, during which the optomechanical
interaction is off $\mu(t)=0$, we assume that a phase shift $\phi=\phi_{0}+\delta\phi$
is applied to the mechanical mode; here $\phi_{0}$ is a known reference
phase that we can adjust and $\delta\phi$ is the (unknown) signal
contribution. During this interval, mechanical thermal decoherence
will continue to occur, resulting in the mechanical operator 
\begin{equation}
\hat{b}^{\dagger}(t_{3})=\hat{b}^{\dagger}(t_{2})e^{-i\phi}+\int_{t_{2}}^{t_{3}}dt'\sqrt{\Gamma}\hat{b}_{\text{in}}^{\dagger}(t'),\label{eq:b-t3}
\end{equation}
cf.\ Eq.~(\ref{eq:mech-evol-sol}) [whether the phase shift $\phi$ is assumed to be applied instantaneously
at $t_{2}$ as here or in some other manner makes no difference as
to the impact of the thermal noise contribution in Eq.~(\ref{eq:b-t3})]. 

The leading edge of the output light field generated during the first
optomechanical interaction $t\in[t_{1},t_{2}]$ impinges on the mechanical
system at the onset of the second interaction $t\in[t_{3},t_{4}]$;
we assume that the looping back of the optical field happens with
a transmission efficiency $\eta_\mathrm{tech}$ leading to
the addition of an additional vacuum noise field $\hat{a}_{\text{vac,i}}$. Hence the input
optical field entering the second optomechanical interaction, i.e.,
for $t\in[t_{3},t_{4}]$, is 
\begin{equation}
\hat{a}_{\text{in}}(t)=\sqrt{\eta_\mathrm{tech}}\hat{a}_{\text{out}}(t-[t_{3}-t_{1}])+\sqrt{1-\eta_\mathrm{tech}}\hat{a}_{\text{vac,i}}(t).\label{eq:a-in-int2}
\end{equation}
The continuous output light field produced by the second interaction
is, according to Eqs.~(\ref{eq:a-out-t}) and (\ref{eq:light-transform-h-in-out})
with $t\in[t_{\text{i}}=t_{3},t_{\text{f}}=t_{4}]$,
\begin{multline}
\hat{a}_{\text{out}}(t)  =ih_{\text{out}}^{(2)}(t)\sqrt{e^{M_{2}}-1}\hat{b}^{\dagger}(t_{3})-\int_{t_{3}}^{t}dt'\left[2\delta(t-t')+2\sinh\left(M_{2}/2\right)h_{\text{out}}^{(2)}(t)h_{\text{in}}^{(2)}(t')\right]\hat{a}_{\text{in}}(t')\\+i\sqrt{\mu(t)}\int_{t_{3}}^{t}dt'e^{\frac{1}{2}M(t,t')}\sqrt{\Gamma}\hat{b}_{\text{in}}^{\dagger}(t'),\label{eq:a-out-t-int2}
\end{multline}
where we have defined the net interaction strength of the second interaction
$M_{2}\equiv M(t_{4},t_{3})$. Substituting the above equations
into Eq.~(\ref{eq:a-out-t-int2}), we can achieve the net scattering
relation linking the final output light field $\hat{a}_{\text{out}}(t\in[t_{3},t_{4}])$
to the various input fields $\hat{a}_{\text{in}}(t'\in[t_{1},t_{2}])$,
$\hat{b}_{\text{in}}^{\dagger}(t'\in[t_{1},t_{4}])$, and $\hat{a}_{\text{vac,i}}(t'\in[t_{3},t_{4}])$
as well as the initial mechanical operator $\hat{b}^{\dagger}(t_{1})$,
\begin{equation}
\hat{a}_{\text{out}}(t)=T_{b^{\dagger}(t_{1})}(t)\hat{b}^{\dagger}(t_{1})+\int_{t_{1}}^{t}dt'\Phi_{b_{\text{in}}^{\dagger}}(t,t')\hat{b}_{\text{in}}^{\dagger}(t')+\int_{t_{3}}^{t}dt'\Phi_{a_{\text{vac,i}}}(t,t')\hat{a}_{\text{vac,i}}(t')+\int_{t_{1}}^{t_{2}}dt'\Phi_{a_{\text{in}}}(t,t')\hat{a}_{\text{in}}(t'),\label{eq:a-out-t-combined}
\end{equation}
where we have defined scattering matrix elements for $t\in[t_{3},t_{4}]$
\begin{multline}
T_{b^{\dagger}(t_{1})}(t)\equiv ih_{\text{out}}^{(2)}(t)\sqrt{e^{M_{2}}-1}e^{-i\phi}e^{\frac{1}{2}M_{1}}\\
-i\int_{t_{3}}^{t}dt'\left[2\delta(t-t')+2\sinh\left(M_{2}/2\right)h_{\text{out}}^{(2)}(t)h_{\text{in}}^{(2)}(t')\right]h_{\text{out}}^{(1)}(t'-[t_{3}-t_{1}])\sqrt{\eta_\mathrm{tech}}\sqrt{e^{M_{1}}-1},\label{eq:T-b-t1-gen}
\end{multline}
\begin{equation}
\Phi_{b_{\text{in}}^{\dagger}}(t,t')\equiv i\sqrt{\Gamma}\times\begin{cases}
h_{\text{out}}^{(2)}(t)\sqrt{e^{M_{2}}-1}e^{-\frac{1}{2}M(t',t_{3})} & ,\,t'\in[t_{3},t_{4}]\\
h_{\text{out}}^{(2)}(t)\sqrt{e^{M_{2}}-1} & ,\,t'\in[t_{2},t_{3}]\\
h_{\text{out}}^{(2)}(t)\sqrt{e^{M_{2}}-1}e^{-i\phi}e^{\frac{1}{2}M(t_{2},t')}\\
-\sqrt{\eta_\mathrm{tech}}\sqrt{e^{M_{1}}-1}\Theta(t-[t_{3}-t_{1}]-t')\int_{[t_{3}-t_{1}]+t'}^{t}dt''\\
\quad\times\left[2\delta(t-t'')+2\sinh\left(M_{2}/2\right)h_{\text{out}}^{(2)}(t)h_{\text{in}}^{(2)}(t'')\right]h_{\text{out}}^{(1)}(t''-[t_{3}-t_{1}])e^{-\frac{1}{2}M(t',t_{\text{1}})} & ,\,t'\in[t_{1},t_{2}]
\end{cases}
\end{equation}
\begin{equation}
\Phi_{a_{\text{vac,i}}}(t,t')\equiv-\left[2\delta(t-t')+2\sinh\left(M_{2}/2\right)h_{\text{out}}^{(2)}(t)h_{\text{in}}^{(2)}(t')\right]\sqrt{1-\eta_\mathrm{tech}},
\end{equation}
\begin{multline}
\Phi_{a_{\text{in}}}(t,t')\equiv-h_{\text{out}}^{(2)}(t)\sqrt{e^{M_{2}}-1}e^{-i\phi}e^{\frac{1}{2}M(t_{2},t')}\sqrt{\mu(t')}\\
+\int_{t_{3}}^{t}dt''\left[2\delta(t-t'')+2\sinh\left(M_{2}/2\right)h_{\text{out}}^{(2)}(t)h_{\text{in}}^{(2)}(t'')\right]\sqrt{\eta_\mathrm{tech}}\Theta(t''-[t_{3}-t_{1}]-t')\\
\times\left[2\delta(t''-[t_{3}-t_{1}]-t')+2\sinh\left(M_{1}/2\right)h_{\text{out}}^{(1)}(t''-[t_{3}-t_{1}])h_{\text{in}}^{(1)}(t')\right].
\end{multline}

\subsubsection{Mode-matched interactions}

The above formulas hold for arbitrarily shaped drive envelopes $\mu(t)$
for the two interaction intervals $t\in[t_{1},t_{2}]$ and $t\in[t_{3},t_{4}]$.
We will now assume $\mu(t)$ is chosen so as to match the output mode
of the first interaction to the input mode of the second, $h_{\text{in}}^{(2)}(t)=h_{\text{out}}^{(1)}(t-[t_{3}-t_{1}])$;
moreover, for specificity, we will assume $\mu(t\in[t_{1},t_{2}])=\mu_{1}$,
i.e., the drive envelope is constant during the first interaction.
These assumptions entail
\begin{equation}
h_{\text{in}}^{(2)}(t)=h_{\text{out}}^{(1)}(t-[t_{3}-t_{1}])=\frac{\sqrt{\mu_{1}}e^{\frac{1}{2}\mu_{1}(t-t_{3})}}{\sqrt{e^{M_{1}}-1}}\label{eq:h-match}
\end{equation}
and, using Eqs.~(\ref{eq:M-from-h-in}), (\ref{eq:mu-from-h-in})
and (\ref{eq:h-out-j}), for $t\in[t_{3},t_{4}]$
\begin{equation}
M(t,t_{3})  =-\ln\left(1-\frac{1-e^{-M_{2}}}{e^{M_{1}}-1}(e^{\mu_{1}(t-t_{3})}-1)\right),
\end{equation}
\begin{equation}
\mu(t)=\frac{\mu_{1}}{\frac{e^{M_{1}}-e^{-M_{2}}}{1-e^{-M_{2}}}e^{-\mu_{1}(t-t_{3})}-1},
\end{equation}
\begin{equation}
h_{\text{out}}^{(2)}(t)  =\frac{\sqrt{e^{M_{1}}-1}}{e^{M_{2}/2}-e^{-M_{2}/2}}\frac{\sqrt{\mu_{1}}e^{-\mu_{1}(t-t_{3})/2}}{\frac{e^{M_{1}}-e^{-M_{2}}}{1-e^{-M_{2}}}e^{-\mu_{1}(t-t_{3})}-1}.
\end{equation}

The transfer function of the mechanical initial condition, Eq.~(\ref{eq:T-b-t1-gen}),
simplifies as
\begin{align}
T_{b^{\dagger}(t_{1})}(t) & =ih_{\text{out}}^{(2)}(t)\sqrt{e^{M_{2}}-1}e^{-i\phi}e^{\frac{1}{2}M_{1}}-i\int_{t_{3}}^{t}dt'\left[2\delta(t-t')+2\sinh\left(M_{2}/2\right)h_{\text{out}}^{(2)}(t)h_{\text{in}}^{(2)}(t')\right]h_{\text{in}}^{(2)}(t')\sqrt{\eta_\mathrm{tech}}\sqrt{e^{M_{1}}-1}\nonumber \\
 & =ih_{\text{out}}^{(2)}(t)\left[\sqrt{e^{M_{2}}-1}e^{-i\phi}e^{\frac{1}{2}M_{1}}-e^{M_{2}/2}\sqrt{\eta_\mathrm{tech}}\sqrt{e^{M_{1}}-1}\right],\label{eq:T-b-t1-match}
\end{align}
where we have used the mapping of $h_{\text{in}}^{(2)}(t)$ into $e^{M_{2}/2}h_{\text{out}}^{(2)}(t)$
as given by Eq.~(\ref{eq:h-in-to-out}).

The transfer function for the thermal noise from mechanical decoherence
simplifies to
\begin{equation}
\Phi_{b_{\text{in}}^{\dagger}}(t,t')\equiv i\sqrt{\Gamma}\times\begin{cases}
h_{\text{out}}^{(2)}(t)\sqrt{e^{M_{2}}-1}e^{-\frac{1}{2}M(t',t_{3})} & ,\,t'\in[t_{3},t_{4}]\\
h_{\text{out}}^{(2)}(t)\sqrt{e^{M_{2}}-1} & ,\,t'\in[t_{2},t_{3}]\\
h_{\text{out}}^{(2)}(t)\sqrt{e^{M_{2}}-1}e^{-i\phi}e^{M_{1}/2}e^{-\frac{1}{2}\mu_{1}(t'-t_{1})}\\
+\sqrt{\eta_\mathrm{tech}}\Theta(t-[t_{3}-t_{1}]-t')\bigg[\frac{2\sinh\left(M_{2}/2\right)}{\sqrt{e^{M_{1}}-1}}h_{\text{out}}^{(2)}(t)e^{\frac{1}{2}\mu_{1}(t'-t_{1})}\\
\qquad-\Big(\sqrt{\mu_{1}}e^{\frac{1}{2}\mu_{1}(t-t_{3})}+\frac{2\sinh\left(M_{2}/2\right)}{\sqrt{e^{M_{1}}-1}}h_{\text{out}}^{(2)}(t)e^{\mu_{1}(t-t_{3})}\Big)e^{-\frac{1}{2}\mu_{1}(t'-t_{\text{1}})}\bigg] & ,\,t'\in[t_{1},t_{2}]
\end{cases}\label{eq:Phi-binDag-match}
\end{equation}
where we have used Eq.~(\ref{eq:h-match}) and the result
\begin{equation}
\int_{[t_{3}-t_{1}]+t'}^{t}dt''h_{\text{out}}^{(1)2}(t''-[t_{3}-t_{1}])=\int_{[t_{3}-t_{1}]+t'}^{t}dt''\frac{\mu_{1}e^{\mu_{1}(t''-t_{3})}}{e^{M_{1}}-1}=\frac{e^{\mu_{1}(t-t_{3})}-e^{\mu_{1}(t'-t_{1})}}{e^{M_{1}}-1}.
\end{equation}

The transfer functions (\ref{eq:T-b-t1-match}) and (\ref{eq:Phi-binDag-match})
suffice to calculate the phase sensitivity since the transfer functions
of the vacuum input fields $\hat{a}_{\text{in}}(t)$ and $\hat{a}_{\text{vac,i}}(t)$
can be eliminated as we will see in SM Sec.~\ref{app:sensitivity}.

\subsection{Photon counting of transduced multimode reservoirs}\label{app:phot-count-reservoirs}

\subsubsection{Problem formulation}

The goal of the present calculation is to calculate the statistical
moments of the photon counting operator 
\begin{equation}
\hat{N}\equiv\int_{0}^{\tau}dt\hat{a}_{\text{out}}^{\dagger}(t)\hat{a}_{\text{out}}(t)\label{eq:N-hat-def}
\end{equation}
for the detection interval $t\in[0,\tau]$, where $\hat{a}_{\text{out}}(t)$
is a continuous field amplitude operator with commutation relation
$[\hat{a}_{\text{out}}(t),\hat{a}_{\text{out}}^{\dagger}(t')]=\delta(t-t')$.
We remark that for a time-resolved photon counting record, one could
choose to apply a (generally non-flat) counting envelope to define
the generalized counting operator $\hat{N}_{c}\equiv\int_{0}^{\tau}dt\,c(t)\hat{a}_{\text{out}}^{\dagger}(t)\hat{a}_{\text{out}}(t)$;
this option will not be explored here for simplicity {[}note that
such $c(t)$ cannot project out temporal modes of the field $\hat{a}_{\text{out}}(t)${]}. 

We will assume the particular form for the output field
\begin{equation}
\hat{a}_{\text{out}}(t)=\sum_{j}\int_{t_{0}}^{t}dt'\Phi_{j}(t,t')\hat{d}_{\text{in},j}(t'),\label{eq:scattering-relation}
\end{equation}
where $t_{0}\leq0$, i.e., the output field $\hat{a}_{\text{out}}(t)$,
defined on the interval $t\in[0,\tau]$, may include reservoir contributions
from the (potentially) larger interval $[t_{0},\tau]$. The index
$j$ runs over a set of input reservoirs described by continuously
defined operators $\hat{d}_{\text{in},j}(t')$ assumed to have commutation
relations
\begin{equation}
[\hat{d}_{\text{in},j}(t'),\hat{d}_{\text{in},j'}^{\dagger}(t'')]=\pm_{j}\delta_{j,j'}\delta(t'-t''),\label{eq:d-in-commutator}
\end{equation}
with all other commutators being zero; here, $\pm_{j}$ is a sign
that is allowed to depend on $j$. Moreover, we will assume the input
states for all reservoirs $j$ to be mutually uncorrelated white noise,
\begin{equation}
\langle\hat{d}_{\text{in},j}^{\dagger}(t')\hat{d}_{\text{in},j'}(t'')\rangle=\delta_{j,j'}D_{j}\delta(t'-t''),\label{eq:d-in-correl}
\end{equation}
parametrized by the occupation coefficient $D_{j}$.

From a formal viewpoint, the time variables $t$ and $t'$, indexing
the output and input spaces respectively, could be regarded as abstract
indices pertaining to separate spaces of operators. Here, however,
we have chosen to incorporate the property of causality in the scattering
relation, Eq.~(\ref{eq:scattering-relation}), by constraining the
upper limit of the integration interval to $t$ rather than $\tau$.

The derivation of the statistical moments of the counting operator
performed below shows that these can be succinctly expressed via what
we will here refer to as the two-point  correlation functions induced
by each input field $j$,
\begin{equation}
C_{j}(t,\tilde{t})\equiv\int_{t_{0}}^{\min\{t,\tilde{t}\}}dt'\Phi_{j}^{*}(t,t')\Phi_{j}(\tilde{t},t');\label{eq:C-j_def}
\end{equation}
these have the Hermitian symmetry property $C_{j}^{*}(t,\tilde{t})=C_{j}(\tilde{t},t)$.
Physically, this describes the first-order coherence in the output
field between times $t$ and $\tilde{t}$ induced by input reservoir
$j$ time-bin mode $t'$, integrated over all causally allowed $t'$.

\paragraph{Identity implied by unitarity}

In the Heisenberg-picture formulation of our scattering problem, Eq.~(\ref{eq:scattering-relation}),
unitarity implies that the scattering matrix, constituted by the elements
$\Phi_{j}(t,t')$, must conserve commutation relations of the fields
involved. Evaluating the commutator $[\hat{a}_{\text{out}}(t),\hat{a}_{\text{out}}^{\dagger}(t')]=\delta(t-t')$
via Eq.~(\ref{eq:scattering-relation}) and the commutation relations
of the input fields, Eq.~(\ref{eq:d-in-commutator}), this leads
to the following resolution of the Dirac delta function,
\begin{align}
\delta(t-\tilde{t}) & =\left[\sum_{j}\int_{t_{0}}^{t}dt'\Phi_{j}(t,t')\hat{d}_{\text{in},j}(t'),\sum_{\tilde{j}}\int_{t_{0}}^{\tilde{t}}d\tilde{t}'\Phi_{\tilde{j}}^{*}(\tilde{t},\tilde{t}')\hat{d}_{\text{in},\tilde{j}}^{\dagger}(\tilde{t}')\right]\nonumber \\
 & =\sum_{j}\pm_{j}\int_{t_{0}}^{t}dt'\int_{t_{0}}^{\tilde{t}}d\tilde{t}'\Phi_{j}(t,t')\Phi_{j}^{*}(\tilde{t},\tilde{t}')\delta(t'-\tilde{t}')\nonumber \\
 & =\sum_{j}\pm_{j}\int_{t_{0}}^{\min\{t,\tilde{t}\}}dt'\int_{t_{0}}^{\min\{t,\tilde{t}\}}d\tilde{t}'\Phi_{j}(t,t')\Phi_{j}^{*}(\tilde{t},\tilde{t}')\delta(t'-\tilde{t}')\nonumber \\
 & =\sum_{j}\pm_{j}\int_{t_{0}}^{\min\{t,\tilde{t}\}}dt'\Phi_{j}(t,t')\Phi_{j}^{*}(\tilde{t},t')\nonumber \\
 & =\sum_{j}\pm_{j}C_{j}(\tilde{t},t);\label{eq:unitarity-identity}
\end{align}
i.e., unitarity constrains this particular sum of two-point correlation
functions, these functions being defined in Eq.~(\ref{eq:C-j_def}).
One important use of the identity~(\ref{eq:unitarity-identity})
is that some of the functions $C_{j}(t,\tilde{t})$, i.e., for one
or more values of $j$, can be eliminated from the final formulas
for the statistical moments.

\paragraph{Embedding discrete modes among the continua}\label{subsec:embed-disc}

The general derivation below will use as its starting point the scattering
relation~(\ref{eq:scattering-relation}), which includes only continuum
input fields. However, a discrete mode can effectively be embedded
among the continua by a procedure to be described in the present subsection;
hence, this trick can be applied to the final results derived below
to extend their applicability to the situation where the set of input
operators {[}i.e., the right-hand side of Eq.~(\ref{eq:scattering-relation}){]}
includes one or more discrete modes. 

We observe that by setting, for a particular $j$, 
\begin{equation}
\Phi_{j}(t,t')=T_{j}(t)h(t'),\label{eq:discrete-Phi-j}
\end{equation}
where $h(t')$ is an arbitrary square-normalized (fictitious temporal
mode) function with support on the interval $t'\in[t_{0},0]$, the
corresponding term in the scattering relation~(\ref{eq:scattering-relation})
can be rewritten as
\begin{equation}
\int_{t_{0}}^{t}dt'\Phi_{j}(t,t')\hat{d}_{\text{in},j}(t')=T_{j}(t)\int_{t_{0}}^{0}dt'h(t')\hat{d}_{\text{in},j}(t')=T_{j}(t)\hat{d}_{j},
\end{equation}
where we have introduced the discrete mode
\begin{equation}
\hat{d}_{j}\equiv\int_{t_{0}}^{0}dt'h(t')\hat{d}_{\text{in},j}(t'),
\end{equation}
whose commutator and statistical properties are linked to those of
the fictitious continuum $\hat{d}_{\text{in},j}(t')$ as
\begin{equation}
[\hat{d}_{j},\hat{d}_{j}^{\dagger}]=\int_{t_{0}}^{0}dt'\int_{t_{0}}^{0}dt''h(t')h^{*}(t'')[\hat{d}_{\text{in},j}(t'),\hat{d}_{\text{in},j}^{\dagger}(t'')]=\pm_{j}\int_{t_{0}}^{0}dt'|h(t')|^{2}=\pm_{j}\label{eq:discrete-commutator}
\end{equation}
and
\begin{equation}
\langle\hat{d}_{j}^{\dagger}\hat{d}_{j}\rangle=\int_{t_{0}}^{0}dt'\int_{t_{0}}^{0}dt''h(t')h^{*}(t'')\left\langle \hat{d}_{\text{in},j}^{\dagger}(t'')\hat{d}_{\text{in},j}(t')\right\rangle =D_{j}\int_{t_{0}}^{0}dt'|h(t')|^{2}=D_{j}.\label{eq:discrete-occupancy}
\end{equation}
In conclusion, Eqs.~(\ref{eq:discrete-commutator}) and (\ref{eq:discrete-occupancy})
show that we should simply assign to our fictitious continuum reservoir
the same commutator sign $\pm_{j}$ and occupancy constant $D_{j}$
as for the discrete mode we wish to embed.

In anticipation of the final formulas derived below, we note that
the above procedure leads to the following two-point correlation function
for the embedded discrete mode [inserting Eq.~(\ref{eq:discrete-Phi-j}]
into Eq.~(\ref{eq:C-j_def})),
\begin{align}
C_{j}(t,\tilde{t}) & =\int_{t_{0}}^{\min\{t,\tilde{t}\}}dt'\Phi_{j}^{*}(t,t')\Phi_{j}(\tilde{t},t')=T_{j}^{*}(t)T_{j}(\tilde{t})\int_{t_{0}}^{0}dt'|h(t')|^{2}=T_{j}^{*}(t)T_{j}(\tilde{t}),\label{eq:C-disc}
\end{align}
which, as should be expected, is independent of the shape of the fictitious
mode function $h(t')$.

\subsubsection{Statistical moments of $\hat{N}$}

\paragraph{First moment $\langle\hat{N}\rangle$}

The photon counting operator~(\ref{eq:N-hat-def}) for the output
field can be reexpressed in terms of the input fields via the scattering
relation~(\ref{eq:scattering-relation}),
\begin{equation}
\hat{N}=\int_{0}^{\tau}dt\hat{a}_{\text{out}}^{\dagger}(t)\hat{a}_{\text{out}}(t)=\int_{0}^{\tau}dt\sum_{j,k}\int_{t_{0}}^{t}dt'\int_{t_{0}}^{t}dt''\Phi_{j}^{*}(t,t')\Phi_{k}(t,t'')\hat{d}_{\text{in},j}^{\dagger}(t')\hat{d}_{\text{in},k}(t'').
\end{equation}
Calculating now the first moment using the assumed input correlation
functions~(\ref{eq:d-in-correl}), we find
\begin{align}
\langle\hat{N}\rangle & =\int_{0}^{\tau}dt\sum_{j}\int_{t_{0}}^{t}dt'\int_{t_{0}}^{t}dt''\Phi_{j}^{*}(t,t')\Phi_{j}(t,t'')\langle\hat{d}_{\text{in},j}^{\dagger}(t')\hat{d}_{\text{in},j}(t'')\rangle\nonumber \\
 & =\int_{0}^{\tau}dt\sum_{j}\int_{t_{0}}^{t}dt'\int_{t_{0}}^{t}dt''\Phi_{j}^{*}(t,t')\Phi_{j}(t,t'')D_{j}\delta(t'-t'')\nonumber \\
 & =\int_{0}^{\tau}dt\sum_{j}\int_{t_{0}}^{t}dt'|\Phi_{j}(t,t')|^{2}D_{j}\nonumber \\
 & =\sum_{j}\int_{0}^{\tau}dtC_{j}(t,t)D_{j},\label{eq:N-expect}
\end{align}
in terms of the two-point correlation functions~(\ref{eq:C-j_def}),
entering only via their diagonal elements $C_{j}(t,t)$, and the coefficients
$D_{j}$, Eq.~(\ref{eq:d-in-correl}), characterizing the thermal
occupancy.

\paragraph{Second moment $\text{Var}[\hat{N}]$}

We proceed by calculating the raw second moment of $\hat{N}$, Eq.~(\ref{eq:N-hat-def}),
again using the scattering relation~(\ref{eq:scattering-relation})
to achieve an expression in terms of the input fields
\begin{align}
\hat{N}^{2}= & \int_{0}^{\tau}dt\sum_{j,k}\int_{t_{0}}^{t}dt'\int_{t_{0}}^{t}dt''\Phi_{j}^{*}(t,t')\Phi_{k}(t,t'')\hat{d}_{\text{in},j}^{\dagger}(t')\hat{d}_{\text{in},k}(t'')\nonumber \\
\times & \int_{0}^{\tau}d\tilde{t}\sum_{\tilde{j},\tilde{k}}\int_{t_{0}}^{\tilde{t}}d\tilde{t}'\int_{t_{0}}^{\tilde{t}}d\tilde{t}''\Phi_{\tilde{j}}^{*}(\tilde{t},\tilde{t}')\Phi_{\tilde{k}}(\tilde{t},\tilde{t}'')\hat{d}_{\text{in},\tilde{j}}^{\dagger}(\tilde{t}')\hat{d}_{\text{in},\tilde{k}}(\tilde{t}'').
\end{align}
Rearranging the order of non-operator factors and taking the expectation
value,
\begin{multline}
\langle\hat{N}^{2}\rangle=\sum_{j,k,\tilde{j},\tilde{k}}\int_{0}^{\tau}dt\int_{t_{0}}^{t}dt'\int_{t_{0}}^{t}dt''\int_{0}^{\tau}d\tilde{t}\int_{t_{0}}^{\tilde{t}}d\tilde{t}'\int_{t_{0}}^{\tilde{t}}d\tilde{t}''\Phi_{j}^{*}(t,t')\Phi_{k}(t,t'')\Phi_{\tilde{j}}^{*}(\tilde{t},\tilde{t}')\Phi_{\tilde{k}}(\tilde{t},\tilde{t}'')\\
\times\langle\hat{d}_{\text{in},j}^{\dagger}(t')\hat{d}_{\text{in},k}(t'')\hat{d}_{\text{in},\tilde{j}}^{\dagger}(\tilde{t}')\hat{d}_{\text{in},\tilde{k}}(\tilde{t}'')\rangle,\label{eq:Nsq-exp1}
\end{multline}
we see that we must identity the combinations of reservoir indices
$(j,k,\tilde{j},\tilde{k})$ that lead to non-zero contributions $\langle\hat{d}_{\text{in},j}^{\dagger}(t')\hat{d}_{\text{in},k}(t'')\hat{d}_{\text{in},\tilde{j}}^{\dagger}(\tilde{t}')\hat{d}_{\text{in},\tilde{k}}(\tilde{t}'')\rangle$;
these are $j=k\neq\tilde{j}=\tilde{k}$, $j=\tilde{k}\neq k=\tilde{j}$,
and $j=k=\tilde{j}=\tilde{k}$. We now evaluate them one by one:\\
$j=k\neq\tilde{j}=\tilde{k}$, 
\begin{equation}
\langle\hat{d}_{\text{in},j}^{\dagger}(t')\hat{d}_{\text{in},j}(t'')\hat{d}_{\text{in},\tilde{j}}^{\dagger}(\tilde{t}')\hat{d}_{\text{in},\tilde{j}}(\tilde{t}'')\rangle=\langle\hat{d}_{\text{in},j}^{\dagger}(t')\hat{d}_{\text{in},j}(t'')\rangle\langle\hat{d}_{\text{in},\tilde{j}}^{\dagger}(\tilde{t}')\hat{d}_{\text{in},\tilde{j}}(\tilde{t}'')\rangle=D_{j}D_{\tilde{j}}\delta(t'-t'')\delta(\tilde{t}'-\tilde{t}'').\label{eq:dquad1}
\end{equation}
$j=\tilde{k}\neq k=\tilde{j}$, 
\begin{align}
\langle\hat{d}_{\text{in},j}^{\dagger}(t')\hat{d}_{\text{in},k}(t'')\hat{d}_{\text{in},k}^{\dagger}(\tilde{t}')\hat{d}_{\text{in},j}(\tilde{t}'')\rangle & =\langle\hat{d}_{\text{in},j}^{\dagger}(t')\hat{d}_{\text{in},j}(\tilde{t}'')\hat{d}_{\text{in},k}(t'')\hat{d}_{\text{in},k}^{\dagger}(\tilde{t}')\rangle\nonumber \\
 & =\langle\hat{d}_{\text{in},j}^{\dagger}(t')\hat{d}_{\text{in},j}(\tilde{t}'')[\hat{d}_{\text{in},k}(t''),\hat{d}_{\text{in},k}^{\dagger}(\tilde{t}')]\rangle+\langle\hat{d}_{\text{in},j}^{\dagger}(t')\hat{d}_{\text{in},j}(\tilde{t}'')\hat{d}_{\text{in},k}^{\dagger}(\tilde{t}')\hat{d}_{\text{in},k}(t'')\rangle\nonumber \\
 & =\pm_{j}D_{j}\delta(t'-\tilde{t}'')\delta(t''-\tilde{t}')+D_{j}D_{k}\delta(t'-\tilde{t}'')\delta(\tilde{t}'-t'')\nonumber \\
 & =D_{j}\left[D_{k}\pm_{k}1\right]\delta(t'-\tilde{t}'')\delta(\tilde{t}'-t'').\label{eq:dquad2}
\end{align}
$j=k=\tilde{j}=\tilde{k}$,
\begin{align}
\langle\hat{d}_{\text{in},j}^{\dagger}(t')\hat{d}_{\text{in},j}(t'')\hat{d}_{\text{in},j}^{\dagger}(\tilde{t}')\hat{d}_{\text{in},j}(\tilde{t}'')\rangle & =\langle\hat{d}_{\text{in},j}^{\dagger}(t')[\hat{d}_{\text{in},j}(t''),\hat{d}_{\text{in},j}^{\dagger}(\tilde{t}')]\hat{d}_{\text{in},j}(\tilde{t}'')\rangle+\langle\hat{d}_{\text{in},j}^{\dagger}(t')\hat{d}_{\text{in},j}^{\dagger}(\tilde{t}')\hat{d}_{\text{in},j}(t'')\hat{d}_{\text{in},j}(\tilde{t}'')\rangle\nonumber \\
 & =\pm_{j}D_{j}\delta(t'-\tilde{t}'')\delta(t''-\tilde{t}')+\langle\hat{d}_{\text{in},j}^{\dagger}(t')\hat{d}_{\text{in},j}^{\dagger}(\tilde{t}')\hat{d}_{\text{in},j}(t'')\hat{d}_{\text{in},j}(\tilde{t}'')\rangle\nonumber \\
 & =D_{j}(D_{j}\pm_{j}1)\delta(t'-\tilde{t}'')\delta(t''-\tilde{t}')+D_{j}^{2}\delta(\tilde{t}'-\tilde{t}'')\delta(t'-t''),\label{eq:dquad3}
\end{align}
where we have assumed that we can ignore the case where all time arguments
are equal $t'=\tilde{t}'=t''=\tilde{t}''$ (since it constitutes a
lower-dimensional subset of the integration volume), 
in which case
\begin{align}
\langle\hat{d}_{\text{in},j}^{\dagger}(t')\hat{d}_{\text{in},j}^{\dagger}(\tilde{t}')\hat{d}_{\text{in},j}(t'')\hat{d}_{\text{in},j}(\tilde{t}'')\rangle & =\sum_{k}P_{k}\langle\{n_{k}(t)\}|\hat{d}_{\text{in},j}^{\dagger}(t')\hat{d}_{\text{in},j}^{\dagger}(\tilde{t}')\hat{d}_{\text{in},j}(t'')\hat{d}_{\text{in},j}(\tilde{t}'')|\{n_{k}(t)\}\rangle\nonumber \\
 & =\sum_{k}P_{k}\left(n_{k}(\tilde{t}'')\delta(\tilde{t}'-\tilde{t}'')\times n_{k}(t'')\delta(t'-t'')+n_{k}(\tilde{t}'')\delta(t'-\tilde{t}'')\times n_{k}(t'')\delta(\tilde{t}'-t'')\right)\nonumber \\
 & =\sum_{k}P_{k}n_{k}(\tilde{t}'')n_{k}(t'')\left(\delta(\tilde{t}'-\tilde{t}'')\delta(t'-t'')+\delta(t'-\tilde{t}'')\delta(\tilde{t}'-t'')\right)\nonumber \\
 & =D_{j}^{2}\left(\delta(\tilde{t}'-\tilde{t}'')\delta(t'-t'')+\delta(t'-\tilde{t}'')\delta(\tilde{t}'-t'')\right).
\end{align}

Inserting these results, Eqs.~(\ref{eq:dquad1})--(\ref{eq:dquad3}),
into Eq.~(\ref{eq:Nsq-exp1}), we find
\begin{align}
\langle\hat{N}^{2}\rangle & =\sum_{j\neq\tilde{j}}\int_{0}^{\tau}dt\int_{t_{0}}^{t}dt'\int_{t_{0}}^{t}dt''\int_{0}^{\tau}d\tilde{t}\int_{t_{0}}^{\tilde{t}}d\tilde{t}'\int_{t_{0}}^{\tilde{t}}d\tilde{t}''\Phi_{j}^{*}(t,t')\Phi_{j}(t,t'')\Phi_{\tilde{j}}^{*}(\tilde{t},\tilde{t}')\Phi_{\tilde{j}}(\tilde{t},\tilde{t}'')D_{j}D_{\tilde{j}}\delta(t'-t'')\delta(\tilde{t}'-\tilde{t}'')\nonumber \\
 & +\sum_{j\neq\tilde{j}}\int_{0}^{\tau}dt\int_{t_{0}}^{t}dt'\int_{t_{0}}^{t}dt''\int_{0}^{\tau}d\tilde{t}\int_{t_{0}}^{\tilde{t}}d\tilde{t}'\int_{t_{0}}^{\tilde{t}}d\tilde{t}''\Phi_{j}^{*}(t,t')\Phi_{\tilde{j}}(t,t'')\Phi_{\tilde{j}}^{*}(\tilde{t},\tilde{t}')\Phi_{j}(\tilde{t},\tilde{t}'')D_{j}\left[D_{\tilde{j}}\pm1\right]\delta(t'-\tilde{t}'')\delta(\tilde{t}'-t'')\nonumber \\
 & +\sum_{j}\int_{0}^{\tau}dt\int_{t_{0}}^{t}dt'\int_{t_{0}}^{t}dt''\int_{0}^{\tau}d\tilde{t}\int_{t_{0}}^{\tilde{t}}d\tilde{t}'\int_{t_{0}}^{\tilde{t}}d\tilde{t}''\Phi_{j}^{*}(t,t')\Phi_{j}(t,t'')\Phi_{j}^{*}(\tilde{t},\tilde{t}')\Phi_{j}(\tilde{t},\tilde{t}'')\nonumber \\
 & \hspace{7cm}\times\left[D_{j}(D_{j}\pm1)\delta(t'-\tilde{t}'')\delta(t''-\tilde{t}')+D_{j}^{2}\delta(\tilde{t}'-\tilde{t}'')\delta(t'-t'')\right]\nonumber \\
 & =\sum_{j,\tilde{j}}\int_{0}^{\tau}dt\int_{0}^{\tau}d\tilde{t}\int_{t_{0}}^{t}dt'\int_{t_{0}}^{\tilde{t}}d\tilde{t}'|\Phi_{j}(t,t')|^{2}|\Phi_{\tilde{j}}(\tilde{t},\tilde{t}')|^{2}D_{j}D_{\tilde{j}}\nonumber \\
 & +\sum_{j,\tilde{j}}\int_{0}^{\tau}dt\int_{0}^{\tau}d\tilde{t}\int_{t_{0}}^{\min\{t,\tilde{t}\}}dt'\int_{t_{0}}^{\min\{t,\tilde{t}\}}d\tilde{t}'\Phi_{j}^{*}(t,t')\Phi_{j}(\tilde{t},t')\Phi_{\tilde{j}}^{*}(\tilde{t},\tilde{t}')\Phi_{\tilde{j}}(t,\tilde{t}')D_{j}(D_{\tilde{j}}\pm1)\nonumber \\
 & =(\sum_{j}\int_{0}^{\tau}dtC_{j}(t,t)D_{j})^{2}+\sum_{j,\tilde{j}}\int_{0}^{\tau}dt\int_{0}^{\tau}d\tilde{t}C_{j}(t,\tilde{t})C_{\tilde{j}}^{*}(t,\tilde{t})D_{j}(D_{\tilde{j}}\pm_{\tilde{j}}1),\label{eq:NsqExp}
\end{align}
again in terms of the two-point correlation functions $C_{j}(t,\tilde{t})$
and the thermal occupancy coefficients $D_{j}$. Hence, in view of
Eq.~(\ref{eq:N-expect}), the central second moment is
\begin{align}
\text{Var}[\hat{N}]=\langle\hat{N}^{2}\rangle-\langle\hat{N}\rangle^{2} & =\sum_{j,\tilde{j}}\int_{0}^{\tau}dt\int_{0}^{\tau}d\tilde{t}C_{j}(t,\tilde{t})C_{\tilde{j}}^{*}(t,\tilde{t})D_{j}(D_{\tilde{j}}\pm_{\tilde{j}}1)\nonumber \\
 & =\sum_{j,\tilde{j}}\int_{0}^{\tau}dt\int_{0}^{\tau}d\tilde{t}\text{Re}\left[C_{j}(t,\tilde{t})C_{\tilde{j}}^{*}(t,\tilde{t})\right]\frac{1}{2}\left[D_{j}(D_{\tilde{j}}\pm_{\tilde{j}}1)+D_{\tilde{j}}(D_{j}\pm_{j}1)\right],\label{eq:VarN}
\end{align}
where in the last line we have achieved a manifestly real expression
by means of the Hermitian property $C_{j}^{*}(t,\tilde{t})=C_{j}(\tilde{t},t)$.

Eqs.~(\ref{eq:N-expect}) and (\ref{eq:VarN}) are the main results of this section. Their evaluation can often be simplified using the identity~(\ref{eq:unitarity-identity}).

\subsection{Multimode sensitivity calculation}\label{app:sensitivity}

We will calculate the sensitivity to the small signal phase $\delta\phi$
relative to a known reference phase $\phi_{0}$ according to the formula
\begin{align}
\Delta \phi & \equiv\left.\frac{\sqrt{\text{Var}[\hat{N}]}}{|d\langle\hat{N}\rangle/d\phi|}\right|_{\phi=\phi_{0}};\label{eq:S-app}
\end{align}
we will evaluate the required moments of the photon count operator
$\hat{N}$, $\langle\hat{N}\rangle$ and $\text{Var}[\hat{N}]$, using
the general results derived in SM Sec.~\ref{app:phot-count-reservoirs},
in particular Eqs.~(\ref{eq:N-expect}) and (\ref{eq:VarN}) that
are phrased in terms of the two-time correlation functions $C_{j}(t,\tilde{t})$,
Eq.~(\ref{eq:C-j_def}), of the various input reservoirs $j$. 

The net scattering relation of the double-pass interaction, Eq.~(\ref{eq:a-out-t-combined}),
obeys the generic form (\ref{eq:scattering-relation}) if we apply
the prescription in Section~\ref{subsec:embed-disc} to accommodate
the discrete mode $\hat{b}^{\dagger}(t_{1})$ representing the initial
oscillator state (assumed to be thermal). The (true) continuum operators
in the scattering relation (\ref{eq:scattering-relation}) are 
\begin{equation}
\hat{d}_{\text{in}}(t')\in\{\hat{b}_{\text{in}}^{\dagger}(t'),\hat{a}_{\text{in}}(t'),\hat{a}_{\text{vac,i}}(t')\}.
\end{equation}
The proportionality constants for the quadratic expectation values
of the input operators are {[}cf.\ Eq.~(\ref{eq:d-in-correl}){]}
\begin{equation}
D_{b_{\text{in}}^{\dagger}}=n_{\text{th}}+1,\;D_{a_{\text{in}}}=0=D_{a_{\mathrm{vac,i}}},\;D_{b^{\dagger}(t_{1})}=n_{0}+1.\label{eq:D-list}
\end{equation}

We now evaluate the first moment~(\ref{eq:N-expect}) using the occupancy
coefficients [Eqs.~\eqref{eq:D-list}]
\begin{align}
\langle\hat{N}\rangle & =(n_{0}+1)\int_{t_{3}}^{t_{4}}dtC_{b^{\dagger}(t_{1})}(t,t)+(n_{\text{th}}+1)\int_{t_{3}}^{t_{4}}dtC_{b_{\mathrm{in}}^{\dagger}}(t,t).\label{eq:Nexpect-spec}
\end{align}

The functions $C_{j}(t,\tilde{t})$ are constrained by unitarity according
to the identity~(\ref{eq:unitarity-identity}), which we will use
to simplify the evaluation of the second statistical moment. Since
we work in the limit where the oscillator thermal decay rate is assumed
to be much smaller than all other rates $\Gamma\ll\mu(t)$ {[}while
the decoherence rate $\sim\Gamma n_{\text{th}}$ may be significant{]},
the corresponding correlation function contribution $C_{b_{\mathrm{in}}^{\dagger}}(t,\tilde{t})\propto\Gamma$
to the identity~(\ref{eq:unitarity-identity}) vanishes,
\begin{equation}
\delta(t-\tilde{t})+C_{b^{\dagger}(t_{1})}(t,\tilde{t})=C_{a_{\mathrm{in}}}(t,\tilde{t})+C_{a_{\mathrm{vac,i}}}(t,\tilde{t});\label{eq:unitarity-identity-spec}
\end{equation}
this particular form is suitable for eliminating the explicit contributions
from the vacuum input fields, $\hat{a}_{\text{in}}(t)$ and $\hat{a}_{\text{vac,i}}(t)$.
As a mathematical identity, Eq.~(\ref{eq:unitarity-identity-spec})
is exact seeing as the functions appearing in it have been derived
in the limit $\Gamma\ll\mu(t)$, namely ignoring its contribution
to $M(t,t')$, Eq.~(\ref{eq:M-def}). We now evaluate the second
moment~(\ref{eq:VarN})
\begin{align}
\text{Var}[\hat{N}]={} & \int_{t_{3}}^{t_{4}}dt\int_{t_{3}}^{t_{4}}d\tilde{t}\bigg[|C_{b^{\dagger}(t_{1})}(t,\tilde{t})|^{2}D_{b^{\dagger}(t_{1})}(D_{b^{\dagger}(t_{1})}-1)+|C_{b_{\mathrm{in}}^{\dagger}}(t,\tilde{t})|^{2}D_{b_{\mathrm{in}}^{\dagger}}(D_{b_{\mathrm{in}}^{\dagger}}-1)\nonumber \\
 & +\sum_{j\neq b^{\dagger}(t_{1})}\text{Re}\left[C_{b^{\dagger}(t_{1})}(t,\tilde{t})C_{j}^{*}(t,\tilde{t})\right]\left[D_{b^{\dagger}(t_{1})}(D_{j}\pm_{j}1)+D_{j}(D_{b^{\dagger}(t_{1})}-1)\right]\nonumber \\
 & +\sum_{j\neq b_{\mathrm{in}}^{\dagger},b^{\dagger}(t_{1})}\text{Re}\left[C_{b_{\mathrm{in}}^{\dagger}}(t,\tilde{t})C_{j}^{*}(t,\tilde{t})\right]\left[D_{b_{\mathrm{in}}^{\dagger}}(D_{j}\pm_{j}1)+D_{j}(D_{b_{\mathrm{in}}^{\dagger}}-1)\right]\bigg]\nonumber \\
={} & \int_{t_{3}}^{t_{4}}dt\int_{t_{3}}^{t_{4}}d\tilde{t}\bigg[|C_{b^{\dagger}(t_{1})}(t,\tilde{t})|^{2}(n_{0}+1)n_{0}+|C_{b_{\mathrm{in}}^{\dagger}}(t,\tilde{t})|^{2}(n_{\text{th}}+1)n_{\text{th}}\nonumber \\
 & +\text{Re}\left[C_{b^{\dagger}(t_{1})}(t,\tilde{t})[C_{a_{\text{in}}}^{*}(t,\tilde{t})+C_{a_{\text{vac}}}^{*}(t,\tilde{t})]\right](n_{0}+1)+\text{Re}\left[C_{b_{\mathrm{in}}^{\dagger}}(t,\tilde{t})[C_{a_{\text{in}}}^{*}(t,\tilde{t})+C_{a_{\text{vac,i}}}^{*}(t,\tilde{t})]\right](n_{\text{th}}+1)\nonumber \\
 & +\text{Re}\left[C_{b^{\dagger}(t_{1})}(t,\tilde{t})C_{b_{\mathrm{in}}^{\dagger}}^{*}(t,\tilde{t})\right]\left((n_{\text{th}}+1)n_{0}+(n_{0}+1)n_{\text{th}}\right)\bigg]\nonumber \\
={} & (n_{\text{th}}+1)\left(\int_{t_{3}}^{t_{4}}dt\int_{t_{3}}^{t_{4}}d\tilde{t}|C_{b_{\mathrm{in}}^{\dagger}}(t,\tilde{t})|^{2}n_{\text{th}}+\int_{t_{3}}^{t_{4}}dt C_{b_{\mathrm{in}}^{\dagger}}(t,t)\right)\nonumber \\
 & +\int_{t_{3}}^{t_{4}}dt C_{b^{\dagger}(t_{1})}(t,t)(n_{0}+1)\left(\int_{t_{3}}^{t_{4}}dt C_{b^{\dagger}(t_{1})}(t,t)(n_{0}+1)+1\right)\nonumber \\
 & +\int_{t_{3}}^{t_{4}}dt\int_{t_{3}}^{t_{4}}d\tilde{t}\text{Re}\left[C_{b^{\dagger}(t_{1})}(t,\tilde{t})C_{b_{\mathrm{in}}^{\dagger}}^{*}(t,\tilde{t})\right](n_{0}+1)(2n_{\text{th}}+1),\label{eq:VarN-spec}
\end{align}
where the last equality follows by using the identity~(\ref{eq:unitarity-identity-spec})
and the following identity for the discrete-mode double integral,
\begin{equation}\label{eq:disc-factorization}
\int_{t_{3}}^{t_{4}}dt\int_{t_{3}}^{t_{4}}d\tilde{t}|C_{b^{\dagger}(t_{1})}(t,\tilde{t})|^{2}=\left(\int_{t_{3}}^{t_{4}}dtC_{b^{\dagger}(t_{1})}(t,t)\right)^{2},
\end{equation}
as follows from Eq.~(\ref{eq:C-disc}). The two-point  correlation
function $C_{b^{\dagger}(t_{1})}(t,\tilde{t})$ is given by applying
Eq.~(\ref{eq:C-disc}) to the transfer function~(\ref{eq:T-b-t1-match}),
whereas $C_{b_{\mathrm{in}}^{\dagger}}(t,\tilde{t})$ is given by
Eq.~(\ref{eq:C-j_def}) applied to the transfer function~(\ref{eq:Phi-binDag-match}).
The sensitivity calculation is then completed by inserting Eqs.~(\ref{eq:Nexpect-spec})
and (\ref{eq:VarN-spec}) into Eq.~(\ref{eq:S-app}).

\subsubsection{Limit of vanishing mechanical decoherence $\Gamma n_\mathrm{th}\rightarrow 0$}\label{app-sec:sens-noMechDecoh}

In the limit of no mechanical decoherence $\Gamma n_\mathrm{th}\rightarrow 0$, the contributions from the corresponding two-time correlation function vanishes $(n_\mathrm{th}+1)C_{b_{\mathrm{in}}^{\dagger}}(t,\tilde{t}) \rightarrow 0$. The above expressions for the first two moments, Eqs.~\eqref{eq:Nexpect-spec} and \eqref{eq:VarN-spec}, thus reduce to 
\begin{equation}\label{eq:Nexpect-spec-noTh}
\langle\hat{N}\rangle=(n_{0}+1)\int_{0}^{\tau}dtC_{b^{\dagger}(t_{1})}(t,t)
\end{equation}
\begin{equation}\label{eq-app:VarN-noMechDecoh2}
\text{Var}[\hat{N}]  =(n_{0}+1)\int_{0}^{\tau}dtC_{b^{\dagger}(t_{1})}(t,t)\left((n_{0}+1)\int_{0}^{\tau}dtC_{b^{\dagger}(t_{1})}(t,t)+1\right) =\langle\hat{N}\rangle(\langle\hat{N}\rangle+1),
\end{equation}
recovering the familiar relation between mean and variance for thermal states. 
In view of this relation, the sensitivity~\eqref{eq:S-app} can be expressed simply in terms of $\langle\hat{N}\rangle$ [Eq.~\eqref{eq:Nexpect-spec-noTh}] and its first derivative
\begin{equation}
\Delta \phi =\left.\frac{\sqrt{\langle\hat{N}\rangle(\langle\hat{N}\rangle+1)}}{|d\langle\hat{N}\rangle/d\phi|}\right|_{\phi=\phi_{0}}.
\label{eq-SM:S-twomode}
\end{equation}
This is consistent with the results of the ``two-mode'' calculation in SM Sec.~\ref{section: analogous}.

\subsection{Influence of mechanical thermal decoherence}\label{SM-sec:thermal-decoh}

In the main text, we evaluated the sensitivity using the full model, which incorporates both thermal noise and optical loss, based on the experimental parameters of the 2D photonic crystal device~\cite{ren_two-dimensional_2020}. Under these conditions, the total number of thermal quanta entering the system, $n_{\mathrm{th}} \Gamma \tau = 0.01$, is negligible.

To evaluate the impact of the thermal bath, we now analyze a regime of stronger thermal decoherence. This is achieved by reducing $\mu$ (or equivalently $C_\mathrm{q}$), so that a fixed $M_1$ corresponds to a longer pulse duration, thereby increasing $n_{\mathrm{th}} \Gamma \tau$. For clarity, we neglect optical loss, assume perfect temporal mode matching as in Eq.~(\ref{eq:mu2}), and set $\tau_{\text{gap}}=0$.

\begin{figure}[H]
\centering
\includegraphics[width=0.5\columnwidth]{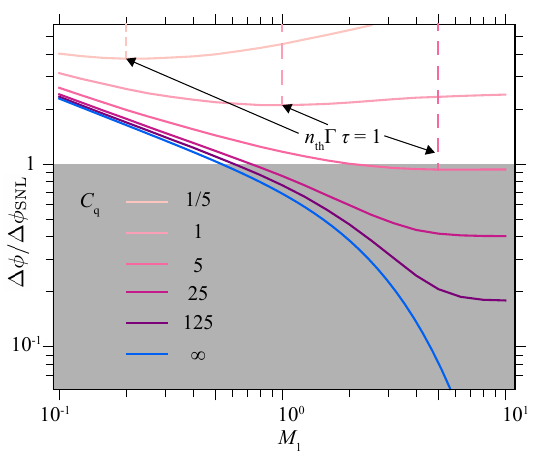}
\caption{\label{fig:1phonon}
SNL-normalized phase sensitivity optimized across $\phi_0$ as a function of $M_1$, evaluated at the optimal $M_2$ (for each $M_1$), for different values of $C_\mathrm{q}$. Parameters: $n_{\mathrm{th}} = 6\times 10^2$, $\Gamma/2\pi = 8$ Hz, and $n_0 = 1$, $\tau_{\mathrm{gap}}=0$. The vertical dashed lines indicate the condition $M_1 /C_{\mathrm{q}} = 1$, which is equivalent to $n_{\mathrm{th}} \Gamma \tau = 1$.
}
\end{figure}

As shown in Fig.~\ref{fig:1phonon}, the phase sensitivity reaches its minimum when $n_{\mathrm{th}} \Gamma \tau \approx 1$, i.e., when $M_1/C_{\mathrm{q}} = 1$. Beyond this point ($M_1 > C_{\mathrm{q}}$), the performance degrades if more than one thermal phonon enters the system. For sufficiently large $C_{\mathrm{q}}$, the sensitivity approaches the ideal case without thermal noise, corresponding to the theoretical curve of Eq.~(\ref{eq:phi-opt-general}) with $\eta_{\mathrm{det}}=1$, $\eta_{12}=1$, and $n_0=1$. This highlights the critical role of $C_{\mathrm{q}}$ in determining the ultimate performance of SU(1,1) interferometry.

\subsection{Fiber loss}
\label{app:fiber_loss}
In the main text, we introduced the overall technical efficiency $\eta_{\mathrm{tech}}$, which accounts for imperfections in storing the first output pulse in the optical fiber. In general, $\eta_{\mathrm{tech}}$ can be decomposed as
\begin{equation}
    \eta_\mathrm{tech} = \eta_\mathrm{in}\,\eta_\mathrm{fiber}\,\eta_\mathrm{out},
\end{equation}
where $\eta_\mathrm{in}$ and $\eta_\mathrm{out}$ denote the input and output coupling efficiencies of the optomechanical cavity, and $\eta_\mathrm{fiber}$ represents the fiber transmission efficiency.  
When evaluating the performance of our scheme, we set $\eta_\mathrm{in}=\eta_\mathrm{out}=1$, and thus assume the fiber loss to be the only significant optical loss channel acting between the two TMS interactions, $\eta_\mathrm{tech}=\eta_\mathrm{fiber}$.  

The fiber transmission efficiency depends on the length $L$ of the fiber used to store the pulse between two interactions, and is given by
\begin{equation}
    \eta_{\text{fiber}} = 10^{-\alpha_{\mathrm{db}} L/10},
\end{equation}
where $\alpha_{\mathrm{db}}$ is the attenuation coefficient with units of dB/km.

\subsection{Phase estimation uncertainty due to parameter fluctuations}
\label{app:error_propagation}

In addition to intrinsic quantum noise, the system is also subject to fluctuations arising in experimental parameters such as laser power variations and other technical noise sources. Here, we parametrize such (classical) fluctuations via fluctuations of the interaction strengths $M_1$ and $M_2$, the internal efficiency $\eta_{12}=\eta_{\mathrm{mode}} \eta_{\mathrm{tech}}$, and the reference phase $\phi_0$. These fluctuations propagate through the system and ultimately contribute to the uncertainty in the measurement outcomes, the impact of which can be quantified through error propagation.
Although these parameter fluctuations do not directly appear in $\text{Var}[\phi]$, they influence the photon detection process, which in turn affects the phase estimation.  

\begin{figure}[h]
	\includegraphics[width=0.49\columnwidth]{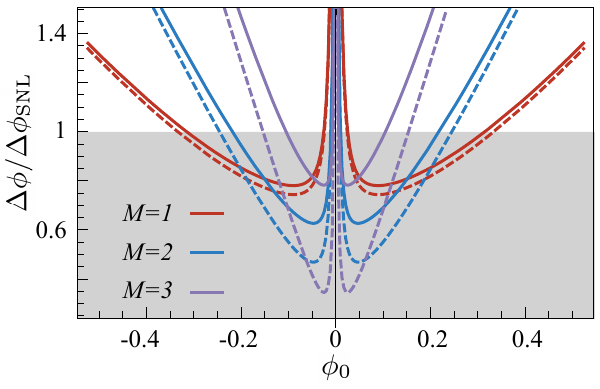}
	\caption{\label{fig:error_0.1} SNL-normalized phase sensitivity as a function of (mean) reference phase $\phi_0$ including fluctuations in (classical) parameters [Eq.~\eqref{eq:exp-fluct}]. Curves represent different (mean) values of $M\equiv M_1 = M_2$ using the experimental parameters of the 2D photonic crystal device considered in the main text (see the near-optimal, orange curve in the inset of Fig.~\ref{fig:ideal_case}(c). Dashed lines represent the ideal case (no error), while solid lines include a 0.1\% relative error (defined as $\sqrt{\mathrm{Var}[P]}/P$  for each parameter $\left.P \in \{M_1, M_2, \eta_{\mathrm{tech}}, \phi_0\}\right.$).
   The grey region lies below the SNL.}
\end{figure}

In particular, $\eta_{\mathrm{mode}}$ as a function of power fluctuations can be set to be one, and its first-order error propagation is zero. This is justified because, in our system, the mode-matching coefficient depends on power fluctuations. At perfect mode matching, the system operates at a local extremum of $\eta_{\mathrm{mode}}$ as a function of power, so the derivative of the mode-matching coefficient with respect to power fluctuations vanishes at this point.
Then, the total variance with error propagation in the photon number measurement can be expressed as 
\begin{align}\label{eq:exp-fluct}
    \text{Var} \!\left[ \hat{N}_{\text{out},2} \right]_{\text{exp}}
    = \text{Var} \!\left[ \hat{N}_{\text{out},2} \right]  & \quad + \left( \frac{\partial \langle\hat{N}_{\text{out},2}\rangle}{\partial M_1} \right)^2 
        \text{Var}[M_1]  + \left( \frac{\partial \langle\hat{N}_{\text{out},2}\rangle}{\partial M_2} \right)^2 
        \text{Var}[M_2] \notag\\
    &\quad + \left( \frac{\partial \langle\hat{N}_{\text{out},2}\rangle}{\partial \eta_{\mathrm{tech}}} \right)^2 
        \text{Var}[\eta_{\mathrm{tech}}] + \left( \frac{\partial \langle\hat{N}_{\text{out},2}\rangle}{\partial \phi_0} \right)^2 
        \text{Var}[\phi_0].
\end{align}
The phase sensitivity that takes into account errors in experimental fluctuations 
\begin{equation}
\Delta \phi _{\mathrm{exp}} = \left.\frac{\sqrt{\text{Var} [ \hat{N}_{\text{out},2} ]_{\text{exp}}}}{|{\partial \langle \hat{N}_{\text{out},2}}\rangle/{\partial \phi}|}\right|_{\phi=\phi_0}.
\end{equation}
Figure~\ref{fig:error_0.1} compares the sensitivity with and without such fluctuations, normalized to the shot-noise limit $\Delta \phi_{\mathrm{SNL}}$. We observe that parameter fluctuations cause a more pronounced degradation at higher squeezing strengths, indicating that the optimal squeezing strength should be re-evaluated when error propagation is taken into account. This behavior is illustrated by the purple and blue curves in Fig.~\ref{fig:error_0.1}, and the observed reductions are reported in the main text.

\end{document}